\documentclass[10pt]{article}
\usepackage[utf8]{inputenc}
\usepackage[margin=1in]{geometry}
\usepackage{natbib}
\usepackage[T1]{fontenc}
\usepackage{amsthm,amssymb,hyperref, bbm, bm, mathrsfs, microtype, mathtools, enumitem}
\usepackage{graphicx, centernot, caption, subcaption, multirow}
\newcommand{\blu}[1]{#1}
\newcommand{\fromrobin}[1]{#1}
\newcommand{\fromrobinn}[1]{#1}
\newcommand{\fromwendy}[1]{#1}

\newcommand{\abs}[1]{\left \lvert #1 \right \rvert} 
\newcommand{\norm}[1]{\left\lVert #1 \right\rVert} 
\newcommand{\E}{\mathbb{E}} 
\newcommand{\Var}[1]{\operatorname{Var} \left( #1 \right)} 
\newcommand{\Corr}[2]{\operatorname{Corr} \left( #1, #2 \right)} 
\newcommand{\inv}{^{-1}}
\newcommand{\mc}{\mathcal}

\newcommand{\CCC}{\operatorname{CCC}}

\newcommand{\sepsj}{\sigma^2_{\bmepsilon_j}}

\newcommand{\tgj}{\tau_{\bmgamma_j}}

\newcommand{\teta}{\tau_{\bmeta}}

\newcommand{\kgj}{k_{\bmgamma_j}}

\newcommand{\keta}{k_{\bmeta}}

\newcommand{\phgj}{\phi_{\bmgamma_j}}

\newcommand{\rOT}{\rho_{12}}

\newcommand{\sgmaj}{\sigma^2_{\bmgamma_j}}

\newcommand{\hrCA}{\hat{\rho}^{\mathrm{CA}}}

\newcommand{\hrReml}{\hat{\rho}^{\mathrm{ReML}}}
\newcommand{\hrOracle}{\hat{\rho}^{\mathrm{oracle}}}
\newcommand{\hrVecchia}{\hat{\rho}^{\mathrm{Vecchia}}}
\newcommand{\hrEblue}{\hat{\rho}^{\mathrm{EBLUE}}}
\newcommand{\hrACE}{\hat{\rho}^{\mathrm{ACE}}}
\newcommand{\hthetaReml}{\hat{\bm{\theta}}^{\mathrm{ReML}}}

\newcommand{\Lr}{\mathcal{L}^{\mathrm{ReML}}}
\newcommand{\cL}{\mathcal{L}}

\newcommand{\bmX}{\bm{X}}
\newcommand{\bmmu}{\bm{\mu}}
\newcommand{\bmeta}{\bm{\eta}}
\newcommand{\bmgamma}{\bm{\gamma}}

\newcommand{\bmepsilon}{\bm{\epsilon}}
\newcommand{\bmZ}{\bm{Z}}
\newcommand{\bmU}{\bm{U}}
\newcommand{\bmA}{\bm{A}}
\newcommand{\bmB}{\bm{B}}
\newcommand{\bmC}{\bm{C}}
\newcommand{\bmJ}{\bm{J}}

\newcommand{\bmV}{\bm{V}}
\newcommand{\bmW}{\bm{W}}
\newcommand{\bmI}{\bm{I}}

\newcommand{\bmG}{\bm{G}}
\newcommand{\bmH}{\bm{H}}

\newcommand{\bmZero}{\bm{0}}

\newcommand{\bmtheta}{\bm{\theta}}

\newcommand{\bmmcI}{\bm{\mathcal{I}}^{\textrm{ReML}}}
\newcommand{\bmmcIvecchia}{\bm{\mathcal{I}}^{\textrm{ML}}}

\newcommand{\bmom}{\bm{\omega}}

\newcommand{\bmLambda}{\bm{\Lambda}}
\newcommand{\bmSigma}{\bm{\Sigma}}
\newcommand{\bmS}{\bm{S}}
\newcommand{\bmR}{\bm{R}}

\newcommand{\bmQ}{\bm{Q}}

\DeclareMathOperator*{\Tr}{Tr}
\DeclareMathOperator*{\arctanh}{arctanh}

\title{A Mixed Model Approach for Estimating Regional Functional Connectivity from Voxel-level BOLD Signals}

\author{Ruobin Liu$^{1}$, Chao Zhang$^{2}$, Chau Tran$^{3}$, Sophie Achard$^{4}$, \\
Wendy Meiring$^{1}$, and Alexander Petersen$^{5}$ \\
\small $^{1}$Department of Statistics and Applied Probability, University of California Santa Barbara, \\
\small Santa Barbara, California, U.S.A. \\
\small $^{2}$Department of Mathematical Sciences, Yeshiva University, New York, New York, U.S.A. \\
\small $^{3}$Department of Statistics, University of California Davis, Davis, California, U.S.A. \\
\small $^{4}$University Grenoble Alpes, CNRS, Inria, Grenoble INP, LJK, F-38000, Grenoble, France \\
\small $^{5}$Department of Statistics, Brigham Young University, Provo, Utah, U.S.A.}
\date{}

\begin{document}
\maketitle






\begin{abstract}
Resting-state brain functional connectivity quantifies the synchrony between activity patterns of different brain regions.  In functional magnetic resonance imaging, each region comprises a set of spatially contiguous voxels at which blood-oxygen-level-dependent signals are acquired. The ubiquitous Correlation of Averages (CA) estimator, and other similar metrics, are computed from spatially aggregated signals within each region, and remain the quantifications of inter-regional connectivity most used by neuroscientists. 
Their popularity is primarily due to computational simplicity despite their demonstrable bias and lack of statistically principled justification. By leveraging linear mixed-effects models, both inter-regional and intra-regional correlation and measurement error can be explicitly modeled as signal variability sources. A novel computational pipeline, focused on subject-level inter-regional correlation parameters of interest, is developed to address the challenges of applying maximum likelihood estimation to such structured, high-dimensional spatiotemporal data. Simulation results \blu{confirm the superiority of the proposed estimator relative to CA in terms of both decreased bias and accurate confidence interval coverage across simulation settings}. The proposed method is \blu{also} applied to construct 
individual human brain networks for subjects from a Human Connectome Project test-retest database. Concordances between inter-regional correlation estimates \blu{demonstrate the potentially substantial scientific benefits of the proposed approach that reliably produces more consistent results than CA for test-retest scans of the same subject.}
\end{abstract}

%



%

\section{Introduction}
\label{sec:intro}


Rapid advancement and increased accessibility of neuroimaging techniques, including functional magnetic resonance imaging (fMRI), have vastly expanded the availability of dynamic brain activity measurements for clinical practice and neuroscience research. \blu{Such data enable advances in modeling and estimating functional brain connectivity, a foundational neuroscience goal due to its importance for studying neurodegenerative diseases and consciousness disorders at both the individual and group levels.}


Network-based approaches are prevalent in functional connectivity studies, modeling each brain as a network where nodes and edges represent brain regions and connections, respectively \citep{fornito2016fundamentals}. \blu{Constructed networks are also frequently used in downstream learning tasks relating connectivity properties to diverse health outcomes.} Following \citet{van2010exploring}, functional connectivity is the dependency of simultaneous neuronal activation patterns of anatomically separated brain regions. In fMRI, activation patterns are collected as blood-oxygen-level-dependent (BOLD) signals over time at numerous spatial locations, or voxels. \blu{Challenges arise in conducting robust and reproducible analyses due to massive data volumes, complex space-time dependencies \citep{achard2023inter, achard2019wavelet}, as well as substantial physiological and measurement noise and other factors \citep{chaimow2018spatial,lohmann2018lisa,park2022clean,termenon2016reliability}.}


In fMRI, the number of voxels is usually high relative to the temporal dimension, leading to computational challenges \fromwendy{in estimating connectivity}. In resting-state analysis, regional homogeneity \citep{jiang2016regional} \blu{quantifies connectivity among nearby voxels (intra-regional connectivity), in contrast to long-range connections between regions (inter-regional connectivity).  It is common in studies of inter-regional connectivity to aggregate voxel measurements within each region before analysis. This aggregation reduces dimension and noise and yields straightforward estimators based on Pearson correlation, termed the Correlation of Averages (CA) estimator, or related measures \citep{fornito2016fundamentals,lbath2024clustering}. However, regional-average approaches generally ignore voxel-level dependence and noise, factors known to bias connectivity estimation \citep{achard2023inter}.}


\blu{Some fMRI studies directly model voxel-level BOLD signals, including mixed models for task-related activation \citep{woolrich2004fully,zhang2014spatio} and models for group-level connectivity analysis \citep{bowman2007spatiotemporal,bowman2008bayesian,zhao2021spatio,chen2024identifying}. However, these approaches are not designed for constructing individual resting-state brain networks and often do not clearly distinguish intra-regional from inter-regional variability \citep{moghimi2022evaluation}. Related frameworks for voxel-level connectivity modeling have primarily focused on improving fixed-effect inference in task fMRI studies \citep{kang2012spatio,kang2017bayesian,castruccio2018scalable,spencer2020joint}.}

\blu{Motivated by these limitations, this work proposes a new estimator for constructing individual-level inter-regional functional connectivity networks with improved reliability and reproducibility. A spatiotemporal mixed-effects model for voxel-level BOLD signals is developed that explicitly characterizes both intra-regional and inter-regional dependence. The proposed approach has two main novelties. First, the estimators for inter-regional connectivity are derived from maximum likelihood rather than moment-based modifications of CA \citep{kang2012spatio,castruccio2018scalable}. Second, an efficient two-stage estimation procedure is developed using Vecchia's likelihood approximation \citep{vecchia_estimation_1988} to overcome the computational burden of Gaussian likelihood methods for massive fMRI data. To the knowledge of the authors, this is the first application of Vecchia's approximation to functional connectivity estimation with voxel-level fMRI data. The proposed estimator is evaluated through simulation studies assessing edge detection performance and through application to repeated scans from the Human Connectome Project (HCP) young adult test-retest database \citep{GLASSER2013105} to assess reproducibility.}


\blu{The remainder of the paper is organized as follows. Section~\ref{sec:mdl} reviews the CA estimator and introduces the proposed mixed model. Section~\ref{sec:est} presents the two-stage estimation procedure and scalable likelihood approximation. Section~\ref{sec:sim} reports simulation studies, while Section~\ref{sec:data} applies the method to HCP test-retest data, demonstrating improved within-individual reproducibility compared with CA. Section~\ref{sec:discussion} concludes with discussion and future directions.}

\section{Background and Model}
\label{sec:mdl}

An individual brain scan consists of time-varying measurements taken across many voxels that are grouped into contiguous regions $\mc{R}_j$, $j = 1,\ldots, J$. 
In \blu{a functional connectivity} network, regions $\mc{R}_j$ are nodes, while connectivity parameters $\rho_{jj'},$ $1 \leq j < j' \leq J,$ determine the edges. In this work, \blu{binary correlation-based networks are considered ($|\rho_{jj'}| \leq 1$), where an edge $(j, j')$ exists if and only if $\rho_{jj'} \neq 0$.}  Alternatively, the correlation value $\rho_{jj'}$ can represent the strength of the edge between these regions in a weighted network.

\blu{
As illustrated in \citet{achard2019wavelet}, wavelets are well suited for long memory time series, with the choice of wavelet level guided by the short memory present in the time series. 
Wavelet-domain representations are especially effective for resting-state fMRI because they isolate scale-specific temporal dependence while reducing high-frequency physiological and scanner noise \citep{PercivalWalden2000,achard2019wavelet}.
}
One could instead map the signals to the frequency domain, but wavelets maintain the ordered nature of the original temporal domain, allowing more structured dependence modeling.  In either case, the transformation is linear, so linear modeling in one space induces a similar model structure in the other. 
\fromwendy{It should be noted that the proposed methods are general in that they may be applied to other scales or (non-wavelet) time-ordered signals, and that preprocessing choices can lead to material differences in the information content of the data and the resulting connectivity estimates.}

Denote by $Y_{jlm}$ the $m$-th wavelet coefficient, $m = 1,\ldots,M,$ of the latent BOLD signal at voxel $v_{jl} \in \mc{R}_j$, $l = 1,\ldots,L_j$, $j=1,\dots,J$.  \blu{Similar to other related works that model voxel-level fMRI data (e.g., \cite{kang2012spatio,castruccio2018scalable}), intra-regional and inter-regional dependence across BOLD signals are induced by defining zero-mean random effects $\eta_{jm}$ and $\gamma_{jlm}$, stationary across $m$, with $Y_{jlm} = \eta_{jm} + \gamma_{jlm}$. Regional effects $\eta_{jm}$ induce inter-regional \fromwendy{dependence} via $\rho_{jj'} = \Corr{\eta_{jm}}{\eta_{j'm}}$; local effects $\gamma_{jlm}$ are correlated only across voxels within the same region $j$, uncorrelated with the regional effects $\eta_{jm}$, and with constant variance within each region. With $\mu_j$ denoting a regional mean level, the observed data are $X_{jlm} = \mu_j + \eta_{jm} + \gamma_{jlm} + \epsilon_{jlm},$ \blu{where $\epsilon_{jlm}$ are uncorrelated zero-mean noise variables with variance $\sepsj$}.  While $\rho_{jj'}$ quantifies correlation between regional effects, another \fromwendy{inter-regional} correlation with potentially greater scientific meaning is $\rho_{jj'}^* = \Corr{Y_{jlm}}{Y_{j'l'm}}$, the pairwise correlation between voxel-level latent BOLD wavelet coefficients.  Letting $\delta_j = \Var{\eta_{jm}}/\Var{Y_{jlm}}$, \fromwendy{these two inter-regional} correlation parameters are related via $\rho_{jj'}^* = \rho_{jj'}\sqrt{\delta_j\delta_{j'}}.$ Importantly, either parameter can be used to assess edge presence or absence.} 

\subsection{Correlation of Averages}
\label{ss: CA}

\blu{In spite of the availability of rich voxel-level information, it is common practice in neuroscience to construct a network using the sample Pearson correlation coefficient between spatially averaged regional signals} $\bar{X}_{jm} = L_j\inv \sum_{l=1}^{L_j} X_{jlm}$.  Letting $\tilde{\mu}_j = M\inv\sum_{m = 1}^M \bar{X}_{jm}$, 
\blu{the Correlation of Averages (CA) estimator is}
\begin{equation}
    \label{eq:rca}
    \hrCA_{jj'} = \frac{\sum_{m=1}^M\left( \bar{X}_{jm} - \tilde{\mu}_j \right)\left( \bar{X}_{j'm} - \tilde{\mu}_{j'} \right)}{\left[\left\{\sum_{m = 1}^M \left(\bar{X}_{jm} - \tilde{\mu}_j\right)^2\right\}\left\{\sum_{m = 1}^M \left(\bar{X}_{j'm} - \tilde{\mu}_{j'}\right)^2\right\}\right]^{1/2}}.
\end{equation}
Although CA is commonly used, only recently has rigorous attention been given to its estimand.  
In cases where a theoretical analysis of \fromrobin{$\hrCA_{jj'}$} has been provided, it has typically been done under the hypothetical framework of one observed signal per region (e.g., \cite{afyouni2019effective, azevedo2022deep}), so that averaging across voxels is merely viewed as a preprocessing step without assessing the additional uncertainty that it induces.  However, it is not clear what voxel-level model, if any, would be consistent with these analyses.  


\blu{Heuristically, one can argue that CA is intended to target the latent correlation parameter $\rho_{jj'}$ because, if $L_j$ is large and the spatial correlation of $\gamma_{jlm}$ is sufficiently weak, $\bar{X}_{jm} \approx \mu_j + \eta_{jm}$ by the law of large numbers.  However,}  \cite{achard2023inter} showed that
\begin{equation}
    \label{eq:rcaLimit1}
    \Corr{\bar{X}_{jm}}{\bar{X}_{j'm}}
    \blu{= \frac{\rho_{jj'}}{\left\{\left(1 + \frac{1 - \delta_j}{\delta_j}\psi_j + \frac{\beta_j}{\delta_j}\right)\left(1 + \frac{1 - \delta_{j'}}{\delta_{j'}}\psi_{j'} + \frac{\beta_{j'}}{\delta_{j'}}\right)\right\}^{1/2}}},
\end{equation}
where $\psi_j = L_j^{-2} \sum_{l,l' = 1}^{L_j} \Corr{\gamma_{jlm}}{\gamma_{jl'm}}$ is the average \blu{correlation between the local random effects},  and \blu{$\beta_j\inv = \Var{Y_{jlm}}/\left\{\sepsj/L_j\right\}$ is} the signal-to-noise ratio.  Hence, under mild \blu{temporal dependence} assumptions, \fromrobin{$\hrCA_{jj'}$} will converge almost surely to the right-hand side of \eqref{eq:rcaLimit1} as the duration of the scan expands and the number $M$ of wavelet coefficients diverges.  It follows from \eqref{eq:rcaLimit1} that \fromrobin{$\hrCA_{jj'}$} can be extremely biased for \blu{$\rho_{jj'}$} even when $L_j$ is large, particularly when spatial correlation is high ($\psi_j \approx 1$) and $\delta_j$ is small. 
\blu{Various ad-hoc corrections to $\hrCA_{jj'}$ have therefore been proposed \citep{achard2023inter,lbath2024clustering}, generally relying on subdividing regions into smaller sub-regional groups of highly correlated voxels, computing CA between inter-regional pairs of sub-regional averaged signals, then summarizing the collection of resulting CA estimates.  These ad-hoc methods require tuning and lack both theoretical analysis and the principled uncertainty quantification needed for edge determination.} 

\subsection{A BOLD Mixed Effects Model}
\label{ss:mdlst}

Let $\bmX_{jl} = (X_{jl1},\ldots,X_{jlM})^T$, $\bmX_j = (\bmX_{j1}^T,\ldots,\bmX_{jL_j}^T)^T$, and $\bmX = (\bmX_1^T,\ldots,\bmX_J^T)^T \in \mathbb{R}^{N},$ $N = M\sum_{j = 1}^J L_j$.  Let $\bmmu = (\mu_1,\ldots,\mu_J)^T$, $\bmeta_j = (\eta_{j1},\ldots,\eta_{jM})^T$, and  $\bmeta = (\bmeta_1^T,\ldots,\bmeta_J^T)^T.$  Set $\bmgamma_{jl} = (\gamma_{jl1},\ldots,\gamma_{jlM})^T$, $\bmgamma_j = (\bmgamma_{j1}^T,\ldots,\bmgamma_{jL_j}^T)^T$, and $\bmgamma = (\bmgamma_1^T,\ldots,\bmgamma_J^T)^T,$ and define the error vector $\bmepsilon$ using the same ordering.  
Let $\bm{1}_n$ denote the column vector of ones of length $n$, $\bmJ_{n_1,n_2} = \bm{1}_{n_1}\bm{1}_{n_2}^T,$ and let $\bmI_n$ be the $n\times n$ identity matrix. $\bmQ$ is the block diagonal matrix with $\bm{1}_{L_j}$, $j = 1,\ldots,J$, forming the diagonal block and define $\bmZ = \bmQ \otimes \bm{1}_M$ and $\bmU = \bmQ \otimes \bmI_M$, where $\otimes$ is the Kronecker product.  The model for the observed BOLD wavelet coefficients is
\begin{equation}
    \label{eq:mdlXvec}
    \bmX = \bmZ\bmmu + \bmU\bmeta + \bmgamma + \bmepsilon.
\end{equation}

\blu{The covariance structure is parameterized by assuming that} $\bmeta$, $\bmgamma,$ and $\bmepsilon$ are mutually uncorrelated zero-mean vectors. Let $\bmSigma = \Var{\bmepsilon}$ be the diagonal matrix formed by diagonal blocks $\bmSigma_j = \sepsj\bmI_{L_jM}$, $j = 1,\ldots, J$.  \blu{Let $k_{\bmeta_j} > 0$} and set $\bm{S} = \mathrm{diag}(k_{\bmeta_1}^{1/2},\ldots,k_{\bmeta_J}^{1/2}).$ With the inter-regional correlation matrix $\bmR = \{\rho_{jj'}\}_{j,j' = 1}^J$ and an $M \times M$ wavelet covariance matrix $\bmA$, set $\Var{\bmeta} = (\bmS\bmR\bmS )\otimes \bmA.$  Further specify $\Var{\bmgamma} = \bmLambda$, where $\bmLambda$ is a block diagonal matrix with covariance matrices $\bmLambda_j$ of dimension $ML_j$, $j = 1,\ldots,J$, on the diagonal blocks.  
To simplify computation for both estimation and inference, assume a separable structure $\bmLambda_j = \bmC_j \otimes \bmB_j$, where $\bmB_j$ and $\bmC_j$ are the $L_j \times L_j$ spatial correlation and $M \times M$ wavelet covariance matrices, respectively.  Further specifications of \blu{$\bmA, \bmB_j,$ and $\bmC_j$} in the numerical experiments will be given later. Due to the problem motivating the model in \eqref{eq:mdlXvec}, the primary parameters of interest are contained in the correlation matrix $\bmR$, while all other components are viewed as nuisance parameters.  Letting $\bmW = \Var{\bmU \bmeta} = (\bmQ \bmS \bmR \bmS \bmQ^T)\otimes \bmA$, the overall covariance is
$    \bmV = \bmW + \bmLambda + \bmSigma.$

Similar models to \eqref{eq:mdlXvec} have been developed for task \blu{rather than resting-state} data, incorporating \fromrobinn{design} matrices reflecting the timing of stimuli.  Thus, the following comparisons correspond to versions of existing models \fromwendy{after omitting their stimuli}.  \blu{The models of \cite{kang2012spatio} and \cite{castruccio2018scalable} are the most similar to \eqref{eq:mdlXvec}.  \cite{kang2012spatio} fit a model with $m$ representing frequencies rather than wavelets, with signals at different frequencies assumed to be independent, and with voxel-level random effects specified as $\gamma_{jlm} = \mathbf{D}_m\gamma^*_{jl}$.  \cite{castruccio2018scalable} modeled the combined random effects $\eta_{jm} + \gamma_{jlm} + \epsilon_{jlm}$, as a $\operatorname{VAR}(2)$ process, with intra-regional and inter-regional correlations arising from the innovations.  
Other relevant works include a multi-subject joint Bayesian hierarchical model for group-level connectivity \citep{bowman2008bayesian} and Bayesian mixed models with random effects $\eta_{jm}$ and $\gamma_{jlm}$ that do not vary with time \citep{kang2017bayesian,spencer2020joint}.}  \blu{Although \eqref{eq:mdlXvec} is similar in many aspects to existing models, the primary contribution of this work is to develop a novel approach to connectivity estimation}. In particular, the current work places primary importance on the inter-regional correlations $\rho_{jj'}$, whereas previous work has \blu{primarily targeted inference for} task-related effects, with connectivity estimation not as carefully formulated.  These differences will be elucidated further in Section~\ref{ss:altEst}.

\section{Model Estimation}
\label{sec:est}

Model \eqref{eq:mdlXvec} coherently incorporates the inter-regional functional connectivity parameters $\rho_{jj'}$ while allowing for heterogeneous intra-regional correlation structure and noise levels. 
Estimation can be approached in many ways, depending on the assumptions placed on the various effects.  In this paper, the utility of \eqref{eq:mdlXvec} will be demonstrated in the Gaussian setting.  
\blu{Let $K$ and $H$ denote stationary spatial and wavelet covariance kernels, respectively.  For instance, in the numerical experiments, $K(\cdot ; \nu, \phi)$ is the Mat\'ern kernel with smoothness parameter $\nu$ and scale parameter $\phi$, while $H(\cdot ; \tau)$ is the Gaussian kernel with scale parameter $\tau.$}

The matrices $\bmA$, $\bmB_j$, and $\bmC_j$ used in specifying the covariance structure are 
\begin{equation}
    \label{eq:latCovModel}
    \begin{split}
    (\bmA)_{mm'} &= H(|m - m'|;\teta) + \sigma_{\bmeta}^2 \mathbf{1}(m = m'), \\
    (\bmB_j)_{mm'} &= \kgj H(|m - m'|; \tgj) + \sigma_{\bmgamma_j}^2\mathbf{1}(m = m'), \\
    (\bmC_j)_{ll'} &= K(\norm{v_{jl} - v_{jl'}}_2; \nu_j, \phgj),
    \end{split}    
\end{equation}
with $\mathbf{1}(\cdot)$ and $\norm{\cdot}_2$ denoting the indicator function and Euclidean norm, respectively. The new parameters $\kgj,\sigma^2_{\bmgamma_j}, \sigma^2_{\bmeta} > 0$ represent variances, with the latter two corresponding to nugget effects that account for the fact that correlation across wavelet coefficients is generally weak. The other parameters $\teta,\tgj,\phgj > 0,$ $j = 1,\ldots,J$ govern the spatial (at both intra- and inter-regional scales) and wavelet correlations.  While the smoothness parameters $\nu_j$ can, in principle, be estimated, this is notoriously difficult, so these are set to $\nu_j = 5/2$ throughout.

Evaluating the full Gaussian likelihood is computationally prohibitive, \blu{being $O(N^3)$ in \fromrobinn{floating-point operations} and $O(N^2)$ in memory with $N$ data points}. 
\fromrobinn{The typical HCP subject in Section~\ref{sec:data} has roughly $20{,}000$ voxels across the regions of interest with $M = 69$ points in the wavelet domain.}
Therefore, \blu{attention will be restricted} to individual pairs of regions since the correlations $\rho_{jj'}$ are the primary parameters of interest.  \blu{A two-stage estimation approach is proposed}, similar to \cite{kang2012spatio} or \cite{castruccio2018scalable}.  In the first step, data for each region \blu{is isolated} to estimate the covariance parameters associated with the intra-regional spatiotemporal structure; in the second step, each pair of regions \blu{is isolated} to estimate the remaining parameters, including the inter-regional correlations.  

\subsection{Stage 1: Estimating Region-Specific Parameters}\label{ssec:stg1}

In the first step, data for each region $\mc{R}_j$ are used separately to estimate regional parameters
\begin{equation}
    \label{eq:intraPara}
    \bmtheta_j = [\kgj, \sgmaj, \phgj, \tgj]^T.
\end{equation}
All signals within a same region share $\bmeta_j$ as a common signal component, the parameters of which \blu{are ignored} in the first estimation stage by the use of restricted maximum likelihood (ReML).  Writing $\bmU_j = \bm{1}_{L_j}\otimes \bmI_M$, $\bmeta_j^* = \mu_j\bm{1}_{M} + \bmeta_j$, $\bmepsilon_j^* = \bmgamma_j + \bmepsilon_j$, and $\bmV_j = \bmLambda_j + \bmSigma_j,$ the marginal model for the data from region $j$ becomes $\bmX_j = \bmU_j\bmeta_j^* + \bmepsilon_j^*$, where $\bmepsilon_j^* \sim \mathcal{N}(0, \bmV_j)$ and $\bmeta_j^*$ is treated as a fixed effect.  ReML \blu{is used} over maximum likelihood (ML) because the variance components \blu{are of primary interest} and not the fixed effects 
\citep{harville1974bayesian, pinheiro2006mixed}.  
Let $\tilde{\bmV}_j = \bmV_j/\sepsj$ be the scaled covariance matrix as a function of $\tilde{\bmtheta}_j = (\kgj/\sepsj, \sgmaj/\sepsj, \phgj, \tgj)$.  The resulting profiled restricted log likelihood equation is
\begin{equation}
    \label{eq:ProfReLogLikStage1}
    \begin{split}
    l_{R,p}(\tilde{\bmtheta}_j|\bmX_j) &= a_j - \frac{1}{2}\log \det(\tilde{\bmV}_j) - \frac{1}{2}\log \det(\bmU_j^T\tilde{\bmV}_j\inv \bmU_j) \\
    &\hspace{1cm} - \frac{(L_j - 1)M}{2}\log\left\{\left(\bmX_j - \bmU_j\tilde{\bmeta}^*_j\right)^T\tilde{\bmV}_j\inv\left(\bmX_j - \bmU_j\tilde{\bmeta}^*_j\right)\right\},
    \end{split}
\end{equation}
where $a_j$ is a constant depending only on $M$ and $L_j$,  $\tilde{\bmeta}^*_j = (\bmU_j^T\bmV_j\inv\bmU_j)\inv\bmU_j^T\bmV_j\inv\bmX_j$, and the profiled noise variance is
$
\tilde{\sigma}^2_{\bmepsilon_j}(\tilde{\bmtheta}_j) = \left\{(L_j - 1)M\right\}^{-1}\left(\bmX_j - \bmU_j\tilde{\bmeta}^*_j\right)^T\tilde{\bmV}_j\inv\left(\bmX_j - \bmU_j\tilde{\bmeta}^*_j\right).
$
Let $\hat{\bmtheta}_j$ be the estimator of $\bmtheta_j$ obtained by maximizing \eqref{eq:ProfReLogLikStage1}, then transforming back to the original scale using the estimated noise variance.  The profiling approach can be problematic if the variance $\sepsj$ is on a vastly different scale from $\kgj$ and $\sgmaj$.  In such cases, it is advantageous to fit a noiseless model, corresponding to $\sepsj = 0$, which \blu{is also estimated} by ReML, but without profiling.  With a slight abuse of notation, this estimator \blu{is still denoted as} $\hat{\bmtheta}_j$. In Web Appendix B, details are provided for how the choice between the full and noiseless model is made in Stage 1 for each subject and each region in the HCP data set.  Execution of Stage 1 is very fast due to the ability to parallelize across regions and the structure of $\bmLambda_j.$

\subsection{Stage 2: Estimating Global and Inter-Regional Parameters}\label{ssec:stg2}
Without loss of generality, consider the case of $J = 2$ regions in \eqref{eq:mdlXvec}.  The full parameter vector is $\bmom = [\bmtheta^T, \bmtheta_1^T, \bmtheta_2^T, \sigma^2_{\bmepsilon_1}, \sigma^2_{\bmepsilon_2}]^T,$
where $\bmtheta_j$ are as in \eqref{eq:intraPara}, $\sepsj$ are the noise variances and the inter-regional parameters are
$    \bmtheta = [\teta, \keta, \rOT, \sigma^2_{\bmeta}]^T.$ A natural initial approach to estimation in Stage 2 is ReML, with restricted log-likelihood
\begin{equation}
    \label{eq:mdlInterReML}
l_R(\bmom \mid \bmX) 
= l_R(\bmtheta,\bmtheta_1,\bmtheta_2,\sigma^2_{\bmepsilon_1},\sigma^2_{\bmepsilon_2}) = a - \frac{1}{2}\log \det(\bmV) - \frac{1}{2}\log \det(\bmZ \bmV \inv \bmZ^T) \\
- \frac{1}{2}\bmX^T\bmH \bmX,
\end{equation}
where 
$    \bmH = \bmV\inv - \bmV\inv\bmZ(\bmZ^T\bmV\inv\bmZ)\inv\bmZ^T\bmV\inv$ 
is the projection of $\bmV\inv$ onto the orthogonal complement of the column space of $\bmZ$.  To leverage Stage 1 estimates, \blu{define}
\begin{equation}
    \label{eq:Stage2ReMLEst}
    \hthetaReml = \arg \max_{\bmtheta} l_R(\bmtheta, \hat{\bmtheta}_1,\hat{\bmtheta}_2,\hat{\sigma}^2_{\bmepsilon_1}, \hat{\sigma}^2_{\bmepsilon_2}).
\end{equation}
This also covers the noiseless case in which $\hat{\sigma}^2_{\bmepsilon_j} = 0$ for either or both of $j = 1,2.$  \blu{Optimization in \eqref{eq:Stage2ReMLEst} is executed using} the Limited-memory Broyden–Fletcher–Goldfarb–Shanno (L-BFGS) quasi-Newton method \citep{nocedal1980updating}. As demonstrated in Section~\ref{sec:sim}, this approach results in estimates $\hrReml_{12}$ that have excellent statistical performance relative to \fromrobin{$\hrCA_{12}$} using simulated data for $J = 3$ regions. 
However, use of the full likelihood requires computation of the Cholesky factor of $\bmV$ at each iteration and can be extremely taxing when \blu{$L_1$ or $L_2$} is large, as in the data examples of Section~\ref{sec:data}.  Instead, \blu{the data} applications use Vecchia's likelihood approximation \citep{vecchia_estimation_1988, guinness_permutation_2018}, yielding a tractable optimization method based on Fisher scoring to produce approximate maximum likelihood estimates.

\subsubsection{Vecchia's Approximation for Stage 2 Estimation}\label{sssec:vecchia}

Let $p_{\bmom}(\bmX)$ denote the joint distribution of $\bmX$ in \eqref{eq:mdlXvec} for the case of $J = 2$ regions, so that $N = M(L_1 + L_2)$. Define the index set
$\mathcal{I} = \left\{(j, l, m):\, l = 1,\ldots,L_j,\, j = 1,2,\, m = 1,\ldots M\right\}$ and let $\pi\colon \{1,\ldots,N\} \rightarrow \mathcal{I}$
be a bijection representing an ordering of the observations.  For any $i = 1,\ldots,N,$ write $X_{\pi(i)}$ for $X_{j_il_im_i}$, where $\pi(i) = (j_i, l_i, m_i).$
Expand $p_{\bmom}(\bmX)$ using conditional distributions as
\begin{equation}
\label{eq:vecchia_ll}
    p_{\bmom}(\bmX) = p_{\bmom}(X_{\pi(1)})\prod_{i=2}^{N} p_{\bmom}(X_{\pi(i)} \mid X_{\pi(1)}, \dotsc, X_{\pi(i-1)}).
\end{equation}
Vecchia's approximation to $p_{\bmom}$ is to replace the $i$-th conditioning set, $\{1,\ldots,i-1\}$, $i > 1$, with a subset $\mathcal{J}_i \subseteq \{1, \dotsc, i-1\}$.  Specifically, for any given ordering $\pi$ and collection of such index subsets $\mathcal{J} = \{\mathcal{J}_i$: $i = 2,\ldots,N\},$ Vecchia's approximation of \eqref{eq:vecchia_ll} is
\begin{equation}
\label{eq:vecchia_approx}
p_{\bmom}(\bmX) \approx p_{\bmom, \pi, \mathcal{J}}(\bmX) = p_{\bmom}(X_{\pi(1)})\prod_{i=2}^{N} p_{\bmom}(X_{\pi(i)} \mid X_{\pi(j)}, j \in \mathcal{J}_i).
\end{equation}
With small sets $\mc{J}_i$, the approximation greatly reduces computational cost since each component of \eqref{eq:vecchia_approx} involves only a $|\mathcal{J}_i|\times |\mathcal{J}_i|$ covariance matrix and the components may be evaluated in parallel. \blu{The implementation used in the applications} extends the Fisher scoring algorithm of \citet{guinness_permutation_2018} to accommodate the specific covariance structure. 
\fromrobinn{Web Appendix C discusses the choice of ordering $\pi$ and conditioning sets $\mathcal{J}_i$.}
For additional efficiency gains, Stage 1 estimates are fixed during the Stage 2 optimization using Vecchia's approximation.

While this approximation can, in principle, be used for the restricted likelihood, doing so destroys the labels of the data points that reflect their location in space and wavelet ordering.  This information being crucial to a judicious selection of the permutation $\pi$ and conditioning sets $\mathcal{J}_i$, Vecchia's approximation \blu{is therefore used to} target the (unrestricted) likelihood of \eqref{eq:mdlXvec}. As there are only two fixed effects in $\bmmu = [\mu_1, \mu_2]^T$, the drawbacks of using ML instead of ReML are minimal. The accuracy of Vecchia's approximation depends on the choices of the permutation $\pi$ and the conditioning sets $\mathcal{J}$. \blu{The value $|\mc{J}_i| = 100$ was found} to be a good balance between accuracy and efficiency and the approach in \citet{guinness_permutation_2018} was followed for selecting $\pi$ and $\mathcal{J}$.
\fromrobin{
For region pair $(j, j')$, denote the ReML and Vecchia's approximation estimate of $\rho_{jj'}$ as $\hrReml_{jj'}$ and $\hrVecchia_{jj'}$, respectively.
}

\subsection{Asymptotic Inference}\label{sec:asymp}

Another advantage of the proposed model is that the influence of the spatiotemporal dependence between signals can be incorporated into the assessment of estimation uncertainty.  For fMRI data, the appropriate asymptotic regime is that of an expanding time window, corresponding to a diverging number $M$ of wavelet coefficients being observed.  While the voxel locations remain fixed, these still play a crucial role since the spatial design is not a regular lattice and the spatial correlation is only locally and not globally stationary. 

\fromrobin{
Relevant asymptotic properties for the proposed maximum likelihood estimators were established by \citet{mardia1984maximum}.
These results apply to the full parameter vector $\bmom \in \mathbb{R}^{p}$ due to its use in Vecchia's approximation; corresponding results for ReML can be found in Web Appendix D.
}
\fromrobin{
For $J = 2$ regions in \eqref{eq:mdlXvec}, the sample size is $N=M(L_1+L_2)$. 
Denote the negative log-likelihood by $\cL(\bmom)$ and write $\bmV(\bmom) = \bmV$ to emphasize the dependence on $\bmom$. Let $\bmV^{(i)}(\bmom) = \partial \bmV(\bmom) / \partial \omega_i$, where $\omega_i$ is the $i$-th element of $\bmom \in \mathbb{R}^p$ and define
$\bmmcIvecchia_N(\bmom) = \left\{\partial^2\cL(\bmom)/\partial\omega_i \partial \omega_j\right\}_{i,j=1}^p$.
The Fisher information matrix is
$\mathbb{E}_{\bmom} (\bmmcIvecchia_N(\bmom)) = \left\{\Tr\left\{ \bmV\inv \bmV^{(i)} \bmV\inv \bmV^{(j)}\right\}/2\right\}_{i,j = 1}^p$.
Under the regularity conditions of \citet{mardia1984maximum}, the full ML estimator $\hat{\bmom}^{\mathrm{ML}}$ satisfies, as $M \rightarrow \infty$,
\begin{equation}
    \label{eq:fisher_info_vecchia0}
    \left\{\E{\bmmcIvecchia_N(\bmom)}\right\}^{1/2} (\hat{\bmom}^{\mathrm{ML}} - \bmom) \overset{\mathrm{D}}{\rightarrow} \mathbf{N}(\bmZero, \bmI_p).
\end{equation}
This asymptotic distribution will not be exact even in infinite samples, as they do not take into account the two stage nature of the estimation process. Nevertheless, the relevant plug-in estimates of the left-hand side of \eqref{eq:fisher_info_vecchia0} are used to approximate the uncertainty in the Stage 2 estimator. Standard methods are then used to construct approximate confidence intervals; see Web Appendix D for details.
In Section~\ref{sec:confint_sim}, these confidence intervals are shown to perform well in simulations compared to standard inference procedures using the CA estimator.
}

\subsection{Some Alternative Estimators}
\label{ss:altEst}

\blu{Although \cite{kang2012spatio} and \cite{castruccio2018scalable} modeled data in the frequency and time domains, respectively, with slightly different random effect structures, it is simple to adapt their respective connectivity estimation strategies to the current model.  \cite{kang2012spatio} proposed to form residual terms $r_{jlm} = X_{jlm} - \hat{\mu}_j - \hat{\gamma}_{jlm}$ as approximations of $\eta_{jm} + \epsilon_{jlm}.$  For any distinct pairs $(j,l)$ and $(j',l')$, the empirical covariance across $m$ of $r_{jlm}$ and $r_{j'l'm}$, denoted $\hat{c}_{jl, j'l'}$, is a moment-based estimate of $\mathrm{Cov}(\eta_{jm}, \eta_{j'm})$, from which \cite{kang2012spatio} constructed the connectivity estimator
\begin{equation}
    \label{eq:kang}
\frac{(L_jL_{j'})^{-1}\sum_{l = 1}^{L_j}\sum_{l' = 1}^{L_{j'}} \hat{c}_{jl, j'l'}}{\left\{\binom{L_j}{2}^{-1}\sum_{l \neq r} \hat{c}_{jl,jr}\right\}^{1/2}\left\{\binom{L_{j'}}{2}^{-1}\sum_{l' \neq r'} \hat{c}_{j'l',j'r'}\right\}^{1/2}}.
\end{equation}
Similarly, \cite{castruccio2018scalable} proposed to approximate the innovations in their VAR($p$) model by estimating fixed effects, including the autoregressive coefficients, computing residuals, then calculating the ordinary CA estimator on the residuals instead of the raw signals.}

\blu{In the context of model \eqref{eq:mdlXvec}, \eqref{eq:kang} can be adopted as an alternative estimator by using empirical best linear unbiased predictors (EBLUPs) $\hat{\eta}_{jm}^*$ and $\hat{\gamma}_{jlm}$ of $\eta_{jm}^* = \mu_j + \eta_{jm}$ and $\gamma_{jlm}$, respectively, along with $\hat{\mu}_j = M\inv\sum_{m = 1}^M \hat{\eta}_{jm}^*$ to construct the residuals $r_{jlm}$; \fromrobin{this modification is referred to as the ``average of covariances'' estimator (ACE), denoted by $\hrACE_{jj'}$}. Given that wavelet coefficients are nearly uncorrelated across $m$, the method of \cite{castruccio2018scalable} is essentially the same as CA.  As a combination of these two approaches, the CA estimator was also computed on EBLUP signals $\hat{\eta}_{jm}^*$ rather than residual estimates or regionally averaged signals, with the resulting estimate \fromrobin{denoted by $\hrEblue_{jj'}$}}.

\section{Simulation Study}\label{sec:sim}
This section describes the simulation studies that demonstrate the favorable performance of \blu{the proposed} model under different signal strengths and intra-regional correlations.

\subsection{Simulation settings}
\label{sec:sim_settings}
In each setting, $100$ sets of BOLD signals \blu{were generated} from $J=3$ regions with $M=60$ wavelet coefficients. The spatial coordinates used come from a live rat experiment \citep{guillaume2020functional} and contain $L_1 = 41$, $L_2 = 25$, and $L_3 = 77$ voxels. For kernels $H$ and $K$ in \eqref{eq:latCovModel}, the Gaussian kernel $H(u; \tau) = \exp(-\tau^2u^2/2)$ and Mat\'ern-$5/2$ kernel, defined by
$K(d; 5/2, \phi) = \left(1 + \sqrt{5}\phi d + (5/3)\phi^2d^2\right)\exp\left(-\sqrt{5}\phi d\right)$, were used \citep{stein1999interpolation}.

\blu{The performance of estimators $\hrReml_{jj'}$ and $\hrVecchia_{jj'}$ is studied} in comparison to \fromrobin{$\hrCA_{jj'}$, $\hrACE_{jj'}$, and $\hrEblue_{jj'}$ in Section~\ref{ss:altEst}} under varying signal strengths relative to spatiotemporal noise.
The parameters $\delta_j$ and $\psi_j$ introduced in Section~\ref{ss: CA} take the forms
\begin{equation}
\label{eq:delta_and_psi}
\delta_j = \frac{k_{\eta_j}(1+\sigma^2_{\bmeta})}{k_{\eta_j}(1+\sigma^2_{\bmeta}) + (k_{\bmgamma_j} + \sigma^2_{\bmgamma_j})}, \quad
\psi_j = \frac{1}{L_j^2}\sum_{l, l' = 1} K(\norm{v_{jl} - v_{jl'}}; 5/2, \phi_{\bmgamma_j}).
\end{equation}
The consideration of these parameters is motivated by \eqref{eq:rcaLimit1}, in which the average of intra-regional correlations can be expressed as $\alpha_j = \delta_j + (1-\delta_j)\psi_j.$ Large values of $\delta_j$ correspond to strong regional signals relative to the intra-regional spatial covariance.
To cover different signal strengths, for each region, $k_{\eta_j}$ and $\phi_{\bmgamma_j}$ \blu{were chosen} such that $\delta_j\in\{0.1, 0.5, 0.7\}$ and $\psi_j\in\{0.2, 0.5, 0.8\}$, while $\sepsj = \sigma^2_{\bmgamma_j} = \sigma^2_{\bmeta} = 0.1$, $k_{\bmgamma_j} = 2$, $\tau_{\bmgamma_j} = 0.5$, and $\tau_{\eta} = 0.25$ \blu{were} fixed. In each setting, $\mu_1=1$, $\mu_2=10$, $\mu_3 = 20$, 
\blu{$\rho_{12} = 0$}, $\rho_{13} = 0.35$, and $\rho_{23}=0.6$ \blu{were} all fixed.

\subsection{Comparison of estimators}\label{sec:compare_est}
\begin{figure}[b]
  \centering
  \includegraphics[width=\linewidth]{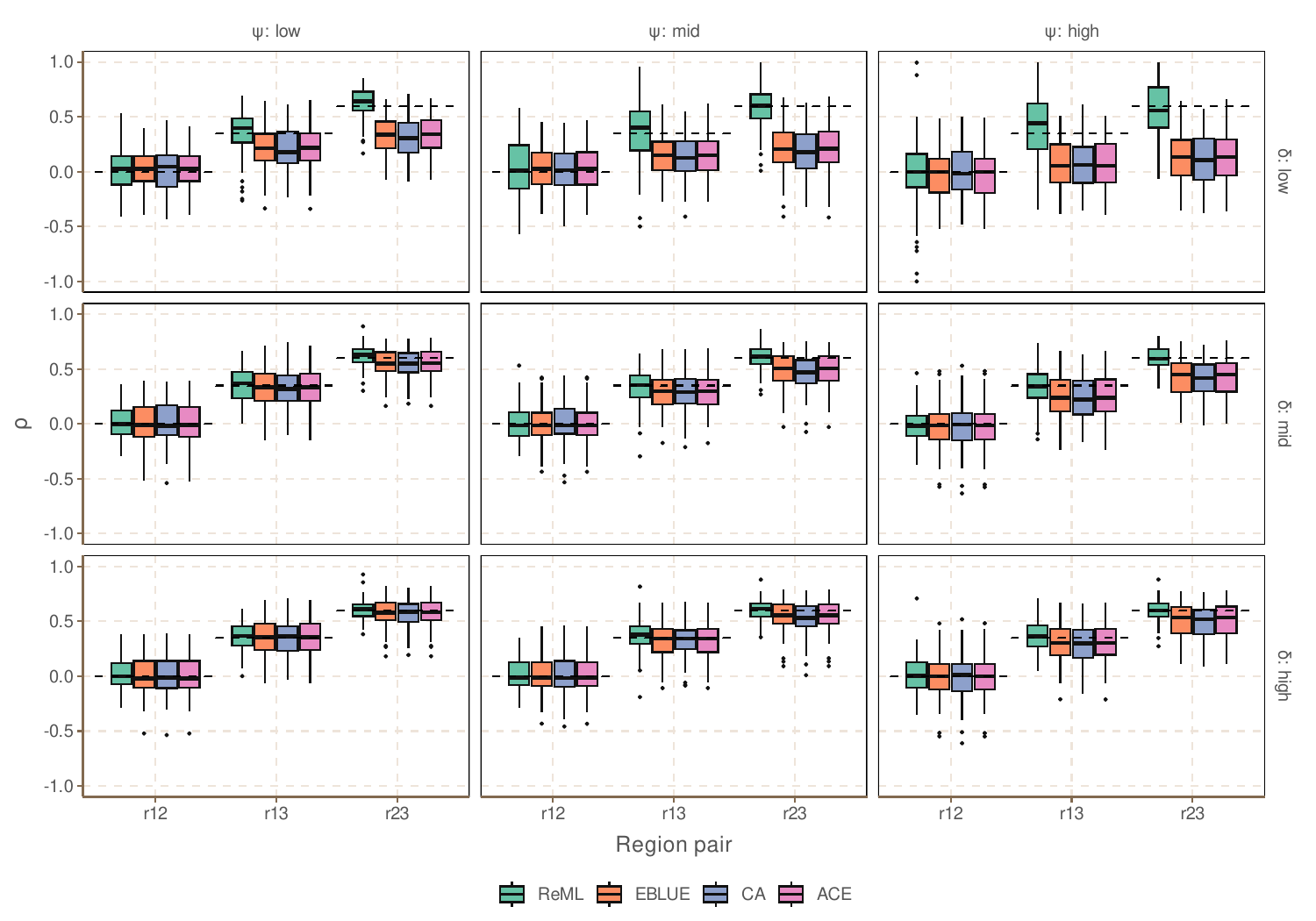}
 \caption{Distribution of $\hrReml_{jj'}$, $\hrEblue_{jj'}$, $\hrCA_{jj'}$, and $\hrACE_{jj'}$ for three region pairs over nine simulation scenarios with $100$ replications each. Rows indicate low ($\delta = 0.1$), medium ($\delta = 0.5$), and high ($\delta = 0.7$) signal strengths while columns indicate low ($\psi = 0.2$), medium ($\psi = 0.5$), and high ($\psi = 0.8)$ intra-regional spatial correlations. The true correlations ($\rho_{12} = 0$, $\rho_{13} = 0.35$, $\rho_{23} = 0.6$) are marked by a dashed line.}
 \label{fig:correctly}
\end{figure}

The performance of the different estimators under each setting is shown in Figure~\ref{fig:correctly} and Table~\ref{tb:simRMSE2}. 
Overall, $\hrReml_{jj'}$ typically has the smallest standard deviation, which can be observed visually in Figure~\ref{fig:correctly}.
The increased spread of $\hrReml_{jj'}$ in the top row reflects the low signal setting ($\delta=0.1$).
Notably, \fromrobin{$\hrCA_{jj'}$, $\hrACE_{jj'}$, and $\hrEblue_{jj'}$} are biased toward $0$ with their medians deviating strongly from the true value when $\rho = 0.6$. Unsurprisingly, the effect of bias seems to be less severe when the signal is strong ($\delta = 0.7$), resulting in better concentrations around the true value for all estimators.  Similarly, for a fixed overall signal strength, \fromrobin{$\hrCA_{jj'}$, $\hrACE_{jj'}$, and $\hrEblue_{jj'}$} are increasingly biased towards $0$ as the average spatial covariance $\psi$ increases. For instance in the high $\psi$ regime, when $\rho = 0.6$, \fromrobin{$\hrCA_{jj'}$, $\hrACE_{jj'}$, and $\hrEblue_{jj'}$} are at least one quartile below the true value, even in the high signal setting. In contrast, $\hrReml_{jj'}$ is still robust with its median staying close to the true inter-regional correlations even though its spread increases. 

The effects of strong spatial covariance on the estimators are tabulated in Table~\ref{tb:simRMSE2}, which shows low ($\psi = 0.2$) and high ($\psi = 0.8$) spatial covariance under a fixed medium signal strength regime ($\delta = 0.5$).
It is seen that a change from low to high spatial covariance leads to a marginal loss in $\hrReml_{jj'}$ but a large loss in \fromrobin{competitor estimators}.
Overall, $\hrReml_{jj'}$ is the most robust estimator across all different simulation settings. Interestingly, $\hrEblue_{jj'}$ performs better than $\hrCA_{jj'}$ for strong spatial covariance and strong intra-regional correlations. This suggests that $\hrEblue_{jj'}$ could be an intermediate estimator for the correlations of the latent signal that can be obtained by running only Stage 1.

\begin{table}[!ht]
\begin{subtable}{\linewidth}
\centering
\begin{tabular}{|c|c|cc|cc|}
\hline
\multirow{2}{*}{$\rho_{jj'}$} & \multirow{2}{*}{Method} & \multicolumn{2}{c|}{$\psi = 0.2$} & \multicolumn{2}{c|}{$\psi = 0.8$} \\
\cline{3-6}
 & & MSE & MAD & MSE & MAD \\
\hline
\multirow{4}{*}{0} & ReML & \textbf{0.020 (0.024)} & \textbf{0.115 (0.079)} & \textbf{0.024 (0.036)} & \textbf{0.119 (0.099)} \\
 & EBLUE & 0.035 (0.044) & 0.153 (0.106) & 0.037 (0.060) & 0.146 (0.125) \\
 & CA & 0.035 (0.044) & 0.154 (0.107) & 0.040 (0.067) & 0.155 (0.129) \\
 & ACE & 0.035 (0.044) & 0.154 (0.106) & 0.037 (0.061) & 0.147 (0.126) \\
\cline{2-6}
\multirow{4}{*}{0.35} & ReML & \textbf{0.021 (0.024)} & \textbf{0.123 (0.076)} & \textbf{0.030 (0.043)} & \textbf{0.137 (0.106)} \\
 & EBLUE & 0.028 (0.037) & 0.136 (0.099) & 0.045 (0.057) & 0.176 (0.119) \\
 & CA & 0.029 (0.037) & 0.140 (0.099) & 0.048 (0.057) & 0.181 (0.123) \\
 & ACE & 0.028 (0.037) & 0.137 (0.099) & 0.045 (0.057) & 0.177 (0.119) \\
\cline{2-6}
\multirow{4}{*}{0.6} & ReML & \textbf{0.009 (0.014)} & \textbf{0.076 (0.058)} & \textbf{0.010 (0.013)} & \textbf{0.084 (0.057)} \\
 & EBLUE & 0.018 (0.030) & 0.106 (0.082) & 0.056 (0.072) & 0.190 (0.143) \\
 & CA & 0.019 (0.032) & 0.106 (0.088) & 0.067 (0.085) & 0.208 (0.156) \\
 & ACE & 0.018 (0.030) & 0.106 (0.082) & 0.056 (0.072) & 0.189 (0.143) \\
\hline
\end{tabular}
\vspace{0.5cm}
\caption{Medium regional signal strength ($\delta=0.5$).}
\end{subtable}
\begin{subtable}{\linewidth}
\vspace{0.5cm}
\centering
\begin{tabular}{|c|c|cc|cc|}
\hline
\multirow{2}{*}{$\rho_{jj'}$} & \multirow{2}{*}{Method} & \multicolumn{2}{c|}{$\delta = 0.1$} & \multicolumn{2}{c|}{$\delta = 0.7$} \\
\cline{3-6}
 & & MSE & MAD & MSE & MAD \\
\hline
\multirow{4}{*}{0} & ReML & 0.064 (0.077) & 0.207 (0.146) & \textbf{0.020 (0.027)} & \textbf{0.112 (0.084)} \\
 & EBLUE & \textbf{0.039 (0.045)} & 0.167 (0.108) & 0.033 (0.044) & 0.141 (0.114) \\
 & CA & 0.042 (0.055) & \textbf{0.160 (0.128)} & 0.036 (0.048) & 0.149 (0.119) \\
 & ACE & 0.041 (0.047) & 0.170 (0.110) & 0.033 (0.044) & 0.141 (0.115) \\
\cline{2-6}
\multirow{4}{*}{0.35} & ReML & \textbf{0.059 (0.105)} & \textbf{0.192 (0.148)} & \textbf{0.023 (0.040)} & \textbf{0.117 (0.095)} \\
 & EBLUE & 0.080 (0.093) & 0.235 (0.158) & 0.026 (0.037) & 0.125 (0.102) \\
 & CA & 0.088 (0.108) & 0.243 (0.170) & 0.025 (0.039) & 0.123 (0.101) \\
 & ACE & 0.080 (0.094) & 0.235 (0.159) & 0.026 (0.037) & 0.125 (0.102) \\
\cline{2-6}
\multirow{4}{*}{0.6} & ReML & \textbf{0.036 (0.058)} & \textbf{0.149 (0.120)} & \textbf{0.009 (0.013)} & \textbf{0.076 (0.058)} \\
 & EBLUE & 0.190 (0.179) & 0.387 (0.201) & 0.021 (0.041) & 0.112 (0.094) \\
 & CA & 0.226 (0.190) & 0.430 (0.205) & 0.025 (0.048) & 0.121 (0.101) \\
 & ACE & 0.189 (0.181) & 0.384 (0.204) & 0.021 (0.041) & 0.112 (0.094) \\
\hline
\end{tabular}
\vspace{0.5cm}
\caption{Medium regional spatial covariance ($\psi=0.5$).}
\end{subtable}
\caption{\label{tb:simRMSE2} Evaluation metrics of the Pearson Correlation of Averages (CA, $\hrCA_{jj'}$), Average of Covariances estimator (ACE, $\hrACE_{jj'}$), Pearson correlation of estimated latent signals (EBLUE, $\hrEblue_{jj'}$), and the proposed ReML estimator (ReML, $\hrReml_{jj'}$) from $100$ simulations and three levels of inter-regional correlation, $\rho_{jj'} = 0, 0.35, 0.6$. (a) For a fixed medium signal strength ($\delta = 0.5$), two levels of intra-regional spatial covariance, $\psi = 0.2$ (weak) and $\psi = 0.8$ (strong) are used. (b) The intra-regional spatial covariance is fixed at a medium level ($\psi = 0.5)$ while the signal strength ranges from $\delta = 0.1$ (weak) to $\delta = 0.7$ (strong).}
\end{table}

\begin{figure}[b]
  \centering
  \includegraphics[width=\linewidth]{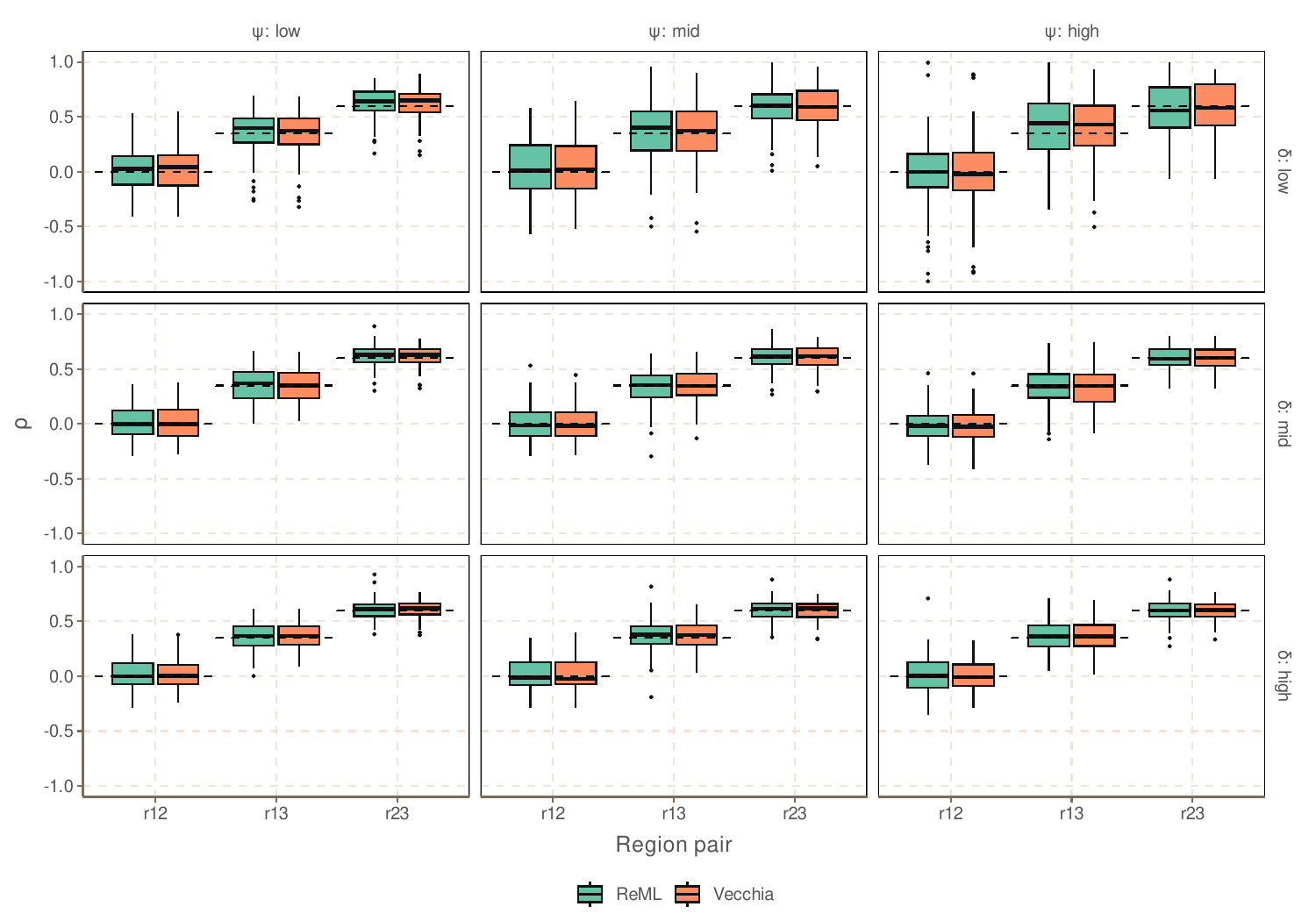}
 \caption{Distribution of $\hrReml_{jj'}$ and $\hrVecchia_{jj'}$ for three region pairs over nine simulation scenarios with $100$ replications each. The simulation settings described in Figure~\ref{fig:correctly} are used with the dashed line marking the true correlation. Results show that using Vecchia's approximation of the likelihood leads to similar performance to that of the full data likelihood.}
 \label{fig:reml_vs_vecchia_plot}
\end{figure}

The next simulation study checks the accuracy of Vecchia's approximation. Figure~\ref{fig:reml_vs_vecchia_plot} compares $\hrReml_{jj'}$ and $\hrVecchia_{jj'}$ under the nine simulation settings mentioned above.
Across all simulation settings, $\hrVecchia_{jj'}$ and $\hrReml_{jj'}$ perform similarly.

To check robustness to model misspecification, the performance of $\hrReml_{jj'}$ \blu{was evaluated} under alternative covariance structures, \blu{including data generation scenarios where $H$ or $K$ are misspecified or where the true covariance of $\gamma_{jlm}$ is not separable across space and wavelet frequencies.  The results generally demonstrate} that $\hrReml_{jj'}$ nevertheless tends to be more accurate than competitors, especially as the signal strength increases.
\blu{The estimator $\hrReml_{jj'}$ was also evaluated} against an oracle estimator where the true Stage 1 coefficients are plugged in and fixed throughout Stage 2, showing that the two-stage procedure leads to minimal loss in performance. See Web Appendix A for these robustness results.

\fromrobinn{
Finally, it is shown that Vecchia's approximation does not result in an estimator that is systematically biased in spatial regimes resembling those of the HCP analysis in Section~\ref{sec:data}.  The estimator $\hrVecchia_{jj'}$ was computed using the same simulation settings in Section~\ref{sec:sim_settings}, except using coordinates from three regions of an arbitrarily selected HCP subject, containing $166$, $233$, and $326$ voxels. See Web Appendix C for full results, which demonstrate that $\hrVecchia_{jj'}$ estimates $\rho_{jj'}$ more accurately than competitors. 
}

\subsection{Coverage of approximate confidence intervals}
\label{sec:confint_sim}
\begin{figure}[b]
  \centering
  \includegraphics[width=\linewidth]{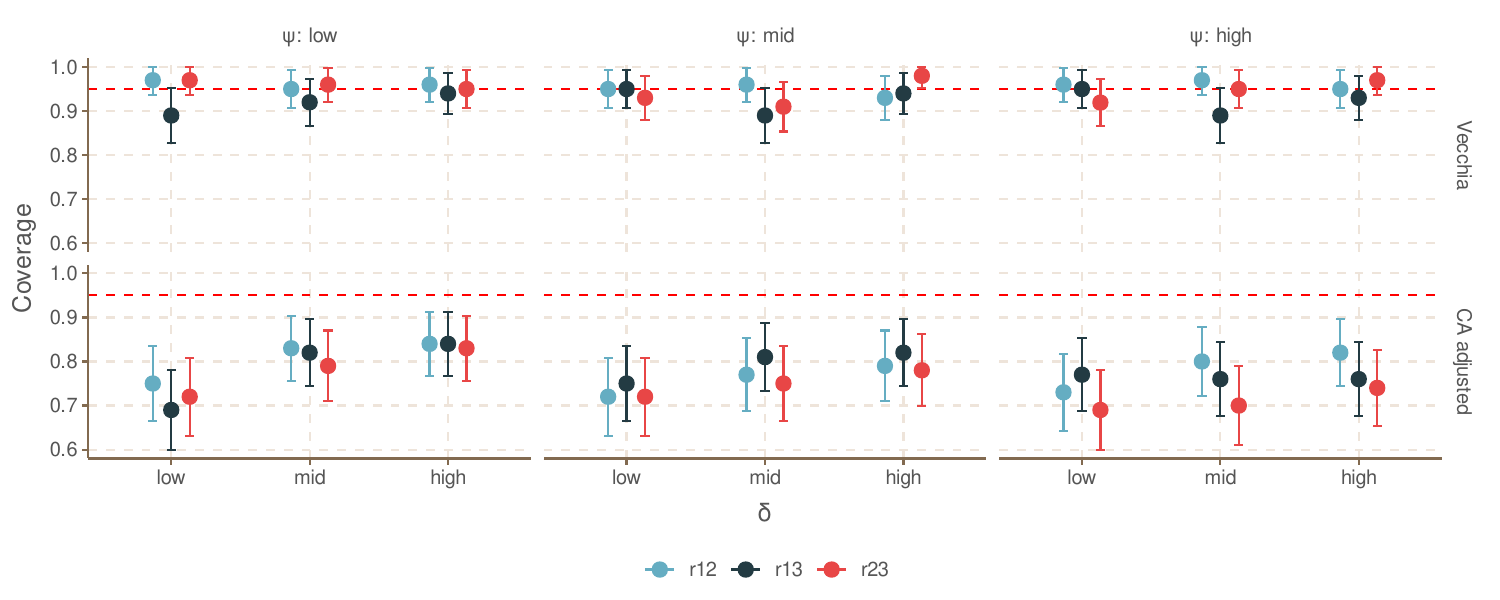}
 \caption{Coverage of $95\%$ asymptotic confidence intervals over $100$ simulations and three levels of inter-regional correlation \fromwendy{for each of Vecchia (top panel) and adjusted CA (lower panel)}. The simulation settings described in Figure~\ref{fig:correctly} are used. The $y$-axis is the proportion of simulations where the $95\%$ confidence interval contains the true parameter, where $\rho_{12} = 0$, $\rho_{13} = 0.35$, and $\rho_{23} = 0.6$, shown as Bernoulli trials.}
 \label{fig:vecchia_95_ci}
\end{figure}

\fromrobin{
Next, the asymptotic development of Section~\ref{sec:asymp} is verified.
Focusing on the HCP application,  Figure~\ref{fig:vecchia_95_ci} shows the proportion of $100$ simulation runs where the $95\%$ confidence interval of $\hrVecchia_{jj'}$ contains $\rho_{jj'}$.
For comparison, confidence intervals based on $\hrCA_{jj'}$ are constructed on the Fisher $Z$ scale using standard error $1/\sqrt{N-3}$, then transformed back to the correlation scale. Since $\hrCA_{jj'}$ does not target $\rho_{jj'}$, $\hrCA_{jj'}$ is first multiplied by the denominator of \eqref{eq:rcaLimit1}, referred to as adjusted CA, which only improves the coverage of the interval. 
Comparing the $\hrVecchia_{jj'}$ intervals with those of the adjusted $\hrCA_{jj'}$ estimator, reasonable coverage is attained by $\hrVecchia_{jj'}$, while the adjusted CA interval systematically fails to contain $\rho_{jj'}$.
Since the true parameter $\rho_{12}$ is zero, the coverage of the confidence intervals for $\hrVecchia_{12}$ checks that the proposed estimator can be used reliably in detecting presence and strength of edges in functional connectivity networks.  In contrast, CA can lead to both an increase in false positives and inaccurate quantification of true edge strength.  See Web Appendix D for further coverage comparisons involving $\hrReml_{jj'}$ and the unadjusted $\hrCA_{jj'}.$
}

\section{Data Applications}\label{sec:data}


\blu{In the main data analysis, the proposed method was applied} to a sample of $42$ subjects out of the 100 considered by \citet{termenon2016reliability} from the young adult HCP test-retest database \citep{GLASSER2013105}, with anatomical regions determined by the AICHA parcellation \citep{JOLIOT2015AICHA}.
\blu{
The Maximum Overlap Discrete Wavelet Transform was used to decompose the BOLD time series of each voxel into multiple temporal scales.
Following previous wavelet-based resting-state fMRI studies, Daubechies orthonormal compactly supported wavelets with filter length $L=8$ were employed \citep{achard2006resilient,termenon2016reliability}. For a repetition time $\mathrm{TR}$, wavelet scale $j$ approximately corresponds to the frequency band $\left[2^{-(j+1)}\mathrm{TR}\inv,2^{-j}\mathrm{TR}\inv\right].$ The HCP resting-state data used here were acquired with $\mathrm{TR}=0.72$s \citep{termenon2016reliability}, so scale 4 corresponds approximately to the frequency range $0.043$--$0.087$ Hz. This range coincides with the low-frequency BOLD fluctuations known to dominate resting-state functional connectivity \citep{biswal1995roi,Cordes2001}. Following \citet{termenon2016reliability} and \citet{achard2006resilient}, scale $4$ wavelets were chosen for these datasets.}

Two scans are available for each subject, referred to as Exam 0 and Exam 1. For all subjects and exams, the $J = 92$ default mode regions \blu{were analyzed}. The voxels per region parameters, $L_j$, ranged approximately from $10$ to $800$ voxels across all subjects.
\fromrobin{For each subject and region pair, $\hrVecchia_{jj'}$ was computed. For a given subject, the median runtime across all region pairs was $2.1$ minutes; see Web Appendix C for details about timing.}
The test-retest setting is widely used to assess reliability of a method in fMRI. In this case, a robust method should give similar connectivity estimates in both exams.
The concordance correlation coefficient (CCC) was used as a similarity metric \citep{lin-1989-concordance}. For random variables $W_j$, $j=0,1$ with mean $\kappa_j$, variance $\varsigma^2_j$, and Pearson correlation $\varrho_{01}$, their population $\CCC$ is $\CCC(W_0, W_1) = 2 \varrho_{01} \varsigma_0 \varsigma_1\left\{\varsigma^2_0 + \varsigma^2_1 + (\kappa_0 - \kappa_1)^2\right\}^{-1}.$
Unlike $\varrho_{01}$, $\CCC(W_0, W_1)$ is sensitive to the location and scale of $W_0$ and $W_1$, so the latter is more appropriate in the current application that assesses competing correlation estimates from a test-retest study. For the HCP data, the sample CCC is computed using plug-ins of the relevant parameters over $\binom{92}{2}$ correlations.

\blu{The estimated connectivity graphs from the proposed mixed model were compared against those from CA} using CCC to assess similarity across test-retest exams.
\fromrobin{
For a fair comparison, the connectivity graphs were constructed using correlation estimates that target the same estimand, for which $\rho_{jj'}^* = \Corr{Y_{jlm}}{Y_{j'l'm}}$ was chosen. While either of $\rho_{jj'}^*$ or $\rho_{jj'}$ can be used to assess the presence of an edge, the raw value of $\rho_{jj'}^*$ is arguably the more scientifically meaningful of the two since it represents the correlation between voxel-level BOLD signals, ignoring white noise. As $\rho_{jj'} = \rho_{jj'}^*/\sqrt{\delta_j\delta_{j'}}$, \eqref{eq:rcaLimit1} becomes
\begin{equation}
\label{eq:rcaLimit2}
\Corr{\bar{X}_{jm}}{\bar{X}_{j'm}} = \frac{\rho_{jj'}^*}{\left[\left\{\delta_j + (1-\delta_j)\psi_j + \beta_j\right\}\left\{\delta_{j'} + (1-\delta_{j'})\psi_{j'} + \beta_{j'}\right\}\right]^{1/2}}.
\end{equation}
Therefore, $\hrCA_{jj'}$ was multiplied by a plug-in estimator
of the denominator in \eqref{eq:rcaLimit2} and $\hrVecchia_{jj'}$ by $\sqrt{\delta_j \delta_{j'}}$ to target $\rho_{jj'}^*$.
}
After this scaling, and some further post-processing as described below, one obtains vectors $\hat{W}_0$ and $\hat{W}_1$ of correlations from Exam 0 and Exam 1, respectively, each with length $\binom{92}{2}$.  Letting $\CCC^\mathrm{MM}$ and $\CCC^\mathrm{CA}$ denote the $\CCC$ computed from the mixed model and CA, \blu{the distribution of these quantities across all subjects can be compared}.

The final vectors of estimated correlations are produced using standard methods. First, a percentage $x$ of the $\binom{92}{2}$ total edges is fixed. Then starting with a $92\times 92$ matrix of estimated correlations produced by the given method, two graphs are produced.  The first graph is constructed by taking the top $x\%$ of edges by the magnitude of the correlations, ignoring any measure of uncertainty; the second graph is constructed by taking the top $x\%$ of edges by magnitude after \blu{setting insignificant correlations to zero.}
Optionally, for each of these graphs, a binarized graph \blu{can be created}, where the non-zero entries are set to one. This results in four graphs for each exam and each method, corresponding to the two thresholding methods and whether or not the graph is binarized.

\begin{figure}[b]
  \centering
  \includegraphics[width=\linewidth]{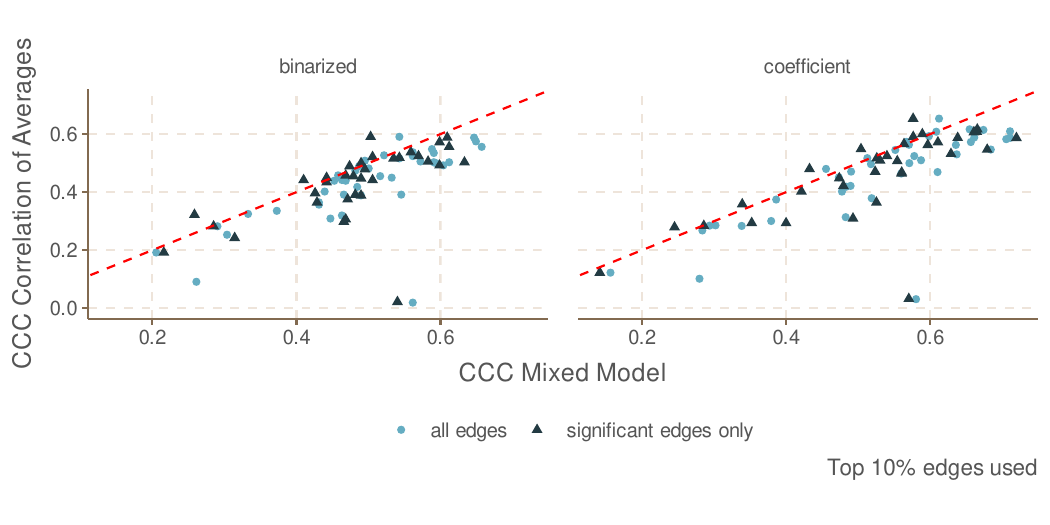}
 \caption{Comparison of concordance correlation coefficients (CCC) between networks estimated from the Correlation of Averages and \blu{the} proposed mixed model to check test–retest reliability across $42$ HCP subjects. Each point represents one subject; circles denote CCC using all edges and triangles denote CCC using only statistically significant edges (Benjamini–Yekutieli FDR control, $q < 0.2$). Panels are faceted by whether graphs were binarized (left) or retained correlation coefficients (right). All graphs were constructed from the top $10\%$ of edges, with exclusions applied when significance filtering yielded fewer edges than required. The dashed red line indicates equality, with points below the line reflecting higher CCC for the mixed model relative to the Correlation of Averages.}
 \label{fig:ccc10}
\end{figure}

\begin{figure}[b]
  \centering
  \includegraphics[width=\linewidth]{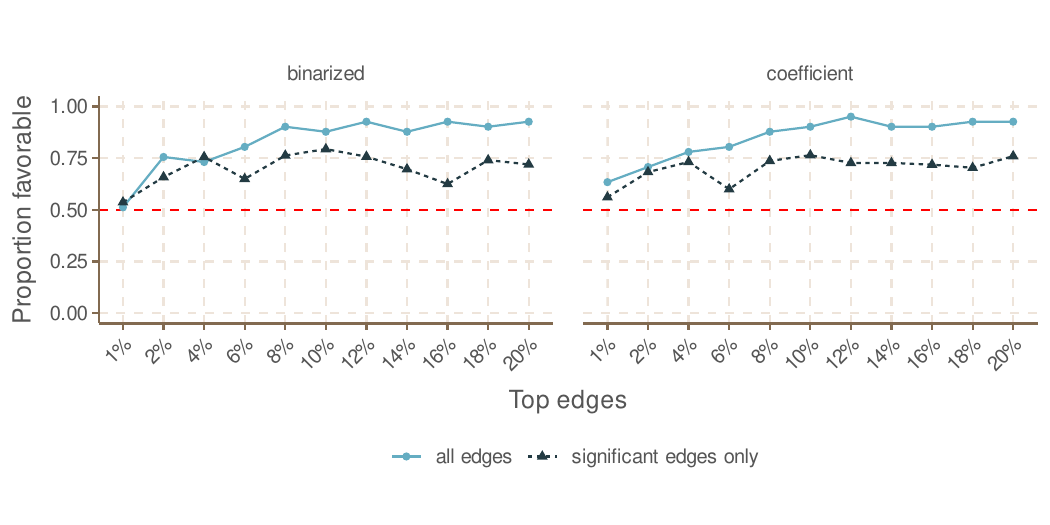}
 \caption{Proportion of the $42$ HCP test-retest subjects with higher concordance correlation coefficient (CCC) for networks constructed from the proposed mixed model relative to those from the Correlation of Averages. The $y$-axis shows the proportion favorable to the mixed model, with the dashed red line marking the $50\%$ reference point (no difference between methods). Results are shown across percentages of top edges used ($x$-axis), comparing graphs constructed from all edges (circles, solid line) versus statistically significant edges only (triangles, dashed line; Benjamini–Yekutieli FDR control, $q < 0.2$). Panels are faceted by whether graphs were binarized (left) or retained correlation coefficients (right).}
 \label{fig:cccprop}
\end{figure}

This procedure is repeated for percentages $x$ going from $1$ to $20$. \blu{The Benjamini-Yekutieli (BY) procedure was used with $q < 0.2$ to perform the significance thresholding \citep{benjamini2001control}.} If a thresholded graph results in fewer than $x\%$ of $\binom{92}{2}$ edges being chosen for either exam, the subject is excluded from the analysis for that choice of $x$.  Figure~\ref{fig:ccc10} plots $\CCC^\mathrm{CA}$ against $\CCC^\mathrm{MM}$ for the top $x\% = 10\%$ of edges. Across all graphs, the majority of points fall under the reference $y=x$ line, suggesting that \blu{the proposed} method achieves greater concordance in each of these graphs.
Figure~\ref{fig:cccprop} plots the proportion of subjects with higher CCC under the proposed method compared to CA for $x\%$ going from $1\%$ to $20\%$. 
\fromrobin{While $q < 0.2$ is a common choice in FDR control, Web Appendix D shows that these results are robust to different BY adjustment thresholds.}

\section{Discussion}
\label{sec:discussion}

In this work, a rigorous statistical framework and computational pipeline \blu{has been developed} for estimating inter-regional resting-state functional connectivity from voxel-level fMRI BOLD signals at the individual level. The widely-used Correlation of Averages (CA) approach ignores intra-regional spatiotemporal dependencies and measurement noise, and yields biased estimates whose asymptotic limits depend on nuisance parameters and the spatial sampling design, rather than the connectivity parameters of scientific interest.  Such biases may impact
studies that use brain connectivity to discriminate between groups or to characterize individuals. 
\blu{The proposed} linear mixed-effects model explicitly accounts for multiple sources of variability, enabling unbiased estimation of inter-regional correlation parameters and improved uncertainty quantification. 
Simulation studies demonstrate that the proposed estimator substantially outperforms the CA across diverse scenarios, with particularly pronounced improvements when intra-regional spatial dependencies are non-negligible \citep{achard2023inter}. 
In \blu{the} proposed approach, rigorous statistical evidence \blu{is demonstrated} in favor of modeling voxel level data \blu{rather than} averaging of voxels within functional connectivity studies, and of utilizing maximum likelihood type estimators in place of more computationally efficient moment-based estimators \citep{kang2012spatio,castruccio2018scalable}.  To alleviate the added computational burden of maximum likelihood estimation, Vecchia's approximation \blu{was introduced} for the first time in the context of functional connectivity estimation.

\blu{The HCP test-retest experiment provides an efficient benchmark for assessing estimator reliability because the subjects are scanned twice.} These datasets have already been used in many studies to compare the efficiency \blu{and reliability of statistical estimators. The empirical analyses provide compelling validation. In the HCP test-retest analysis, networks constructed using the proposed} method exhibited higher concordance between repeated scans.
By moving beyond heuristic averaging procedures to a formal mixed-effects modeling framework with principled inference, this work provides neuroscientists with a method that yields more accurate and reliable estimates of brain network architecture.  \blu{Consequently, there is potential for a more robust scientific understanding of the variability in individual-level connectivity networks among and between different subpopulations, as well as their associations with relevant health outcomes.}

Although designed for estimating functional connectivity, the proposed model could be useful for quantifying dependence among spatially grouped time series in other contexts.  
\fromrobin{For example, similar models as proposed here were used to study climate data \citep{hengl2012spatio, graler2016spatio} and COVID propagation \citep{bartolucci2022spatio}.}

Lastly, the data illustrations have utilized a pre-specified set of regions for functional connectivity analysis \citep{moghimi2022evaluation}. 
This relies on a predefined brain parcellation or atlas to determine regions of interest, so that a common set of functional network nodes is used for different subjects. Recently, data-driven methods for functional connectivity analysis have been proposed \citep{van2010exploring}, thus allowing nodes in functional networks to be subject-specific \citep{cui2020, MICHON2022119589}. It will be interesting to investigate if \blu{the} proposed methods for quantifying functional connectivity can be combined with subject-specific region discovery in order to simultaneously study variability in the spatial distribution of functional connectivity nodes as well as connections between them.




\section*{Acknowledgments}

The authors acknowledge the following facilities for providing computational resources and technical support that have contributed to the results reported in this publication:
\begin{enumerate}
    \item The Office of Research Computing at Brigham Young University. URL: \url{https://rc.byu.edu}.
    \item Use was made of computational facilities purchased with funds from the National Science Foundation (CNS-1725797) and administered by the Center for Scientific Computing (CSC). The CSC is supported by the California NanoSystems Institute and the Materials Research Science and Engineering Center (MRSEC; NSF DMR 2308708) at UC Santa Barbara.
    \item Data were provided [in part] by the Human Connectome Project, WU-Minn Consortium (Principal Investigators: David Van Essen and Kamil Ugurbil; 1U54MH091657) funded by the 16 NIH Institutes and Centers that support the NIH Blueprint for Neuroscience Research; and by the McDonnell Center for Systems Neuroscience at Washington University.

\end{enumerate}

The authors also gratefully acknowledge support by US National Science Foundation's Collaborative Research in Computational Neuroscience program (Award IIS-2135859) and French National Research Agency grants. SA was partly supported by the Agence Nationale de la Recherche under the France 2030 programme, reference ANR-23-IACL-0006. 

\section*{Data Availability}
Data used in this paper to support the findings come from post-processing of the public fMRI young adult test-retest database from the Human Connectome Project, accessible at \url{https://www.humanconnectome.org/study/hcp-young-adult}.
Code to reproduce the results of Section~\ref{sec:sim} is available at \url{https://github.com/roobnloo/qfuncMM-reproducible}.


%
\bibliography{ref.bib}





\clearpage

\setcounter{equation}{0}
\renewcommand{\theequation}{S\arabic{equation}}
\renewcommand{\vec}[1]{\mathbf{#1}}
\renewcommand\thesection{\Alph{section}}
\renewcommand\thesubsection{\thesection.\arabic{subsection}}
\renewcommand{\figurename}{Web Figure}
\renewcommand{\tablename}{Web Table}
\setcounter{section}{0}

\refstepcounter{section}
\section*{Web Appendix A\\Additional simulation results}
\subsection{Robustness to misspecified covariance structure}
In this section we assess through simulations the robustness of our estimator when the model covariance structure is misspecified. Specifically, we consider two alternative temporal covariance kernels and an alternative spatial covariance kernel.

An $\operatorname{AR}(2)$ model has a covariance matrix given by
\[
Q_\mathrm{ar}(|t - t'|) = \frac{\gamma(|t - t'|)}{\gamma(0)},\quad
\gamma(h) = \frac{x^2y^2}{(xy-1)(y-x)} \left(\frac{x^{1-h}}{x^2-1} - \frac{y^{1-h}}{y^2-1}\right),\]
where $x$ and $y$ are parameters such that $\gamma(1)/\gamma(0) = \rho_1$ and $\gamma(2)/\gamma(0) = \rho_2$.
We fix the lag $1$ and lag $2$ autocorrelations to be $\rho_1 = 0.4$ and $\rho_2 = 0.3$, respectively.

Let $\operatorname{fgn}(H)$ denote a fractional Gaussian noise process with Hurst index $H \in (0, 1)$. Then $\operatorname{fgn}(H)$ is governed by the covariance kernel
\[
Q_\mathrm{fgn}(|t - t'|) = \frac{1}{2}\left(|t - t' - 1|^{2H} + |t - t'  +1|^{2H} -2 |t - t'|^{2H}\right),
\]
with $H$ determining the correlation of the process. We use $H=0.9$ in our simulations, which yields highly correlated increments.

\begin{figure}[htbp!]
  \centering
  \includegraphics[width=\linewidth]{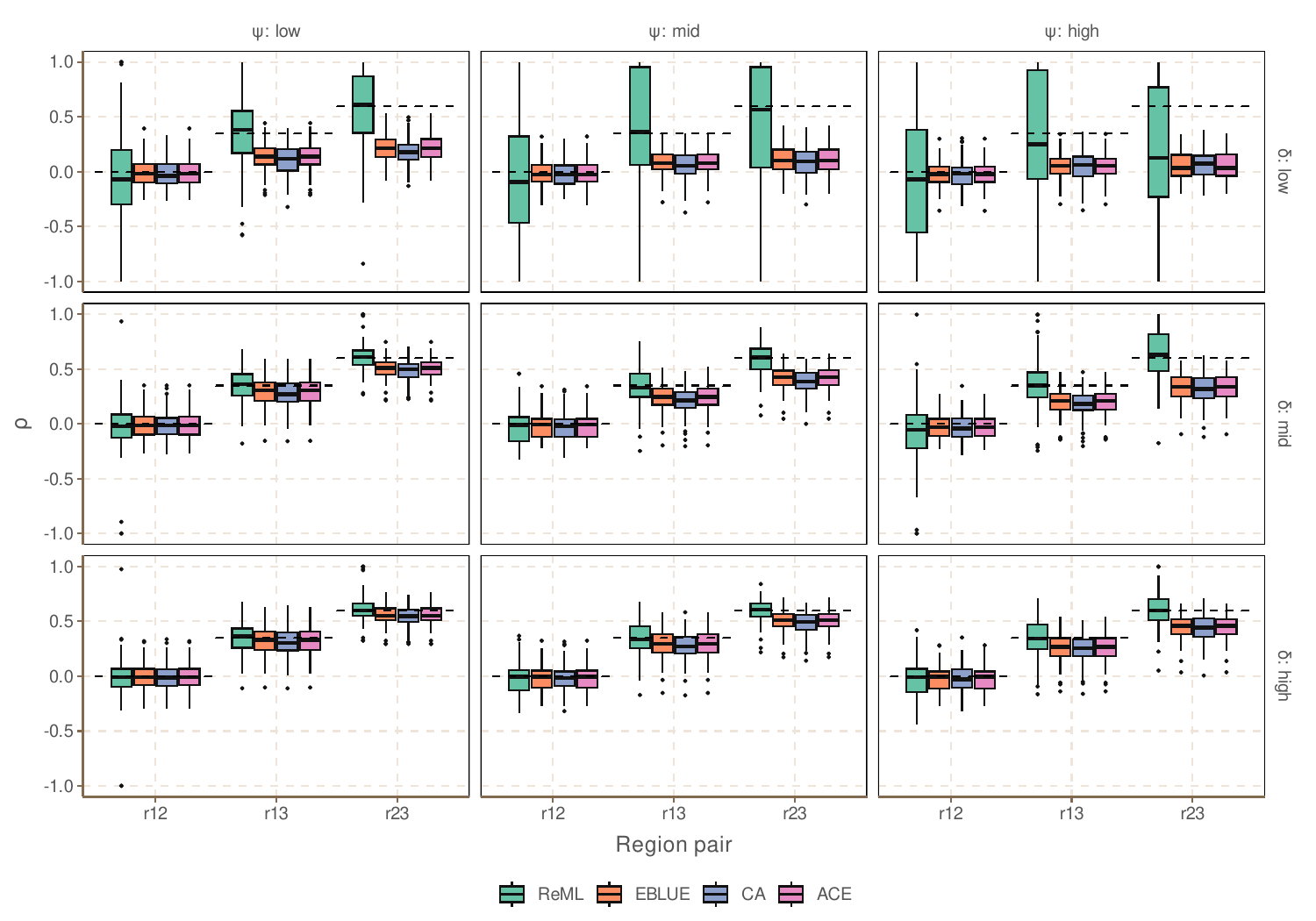}
 \caption{Distribution of $\hrReml_{jj'}$, $\hrEblue_{jj'}$, $\hrCA_{jj'}$, and $\hrACE_{jj'}$ for three region pairs over nine simulation scenarios under a $\operatorname{AR}(2)$ temporal covariance structure
with $100$ replications in each scenario. Rows indicate low ($\delta = 0.1$), medium ($\delta = 0.5$), and high ($\delta = 0.7$) signal strengths while columns indicate low ($\psi = 0.2$), medium ($\psi = 0.5$), and high ($\psi = 0.8)$ intra-regional spatial correlations. The true correlations ($\rho_{12} = 0$, $\rho_{13} = 0.35$, $\rho_{23} = 0.6$) are marked by a dashed line. }
 \label{fig:ar2}
\end{figure}

\begin{figure}[htbp!]
  \centering
  \includegraphics[width=\linewidth]{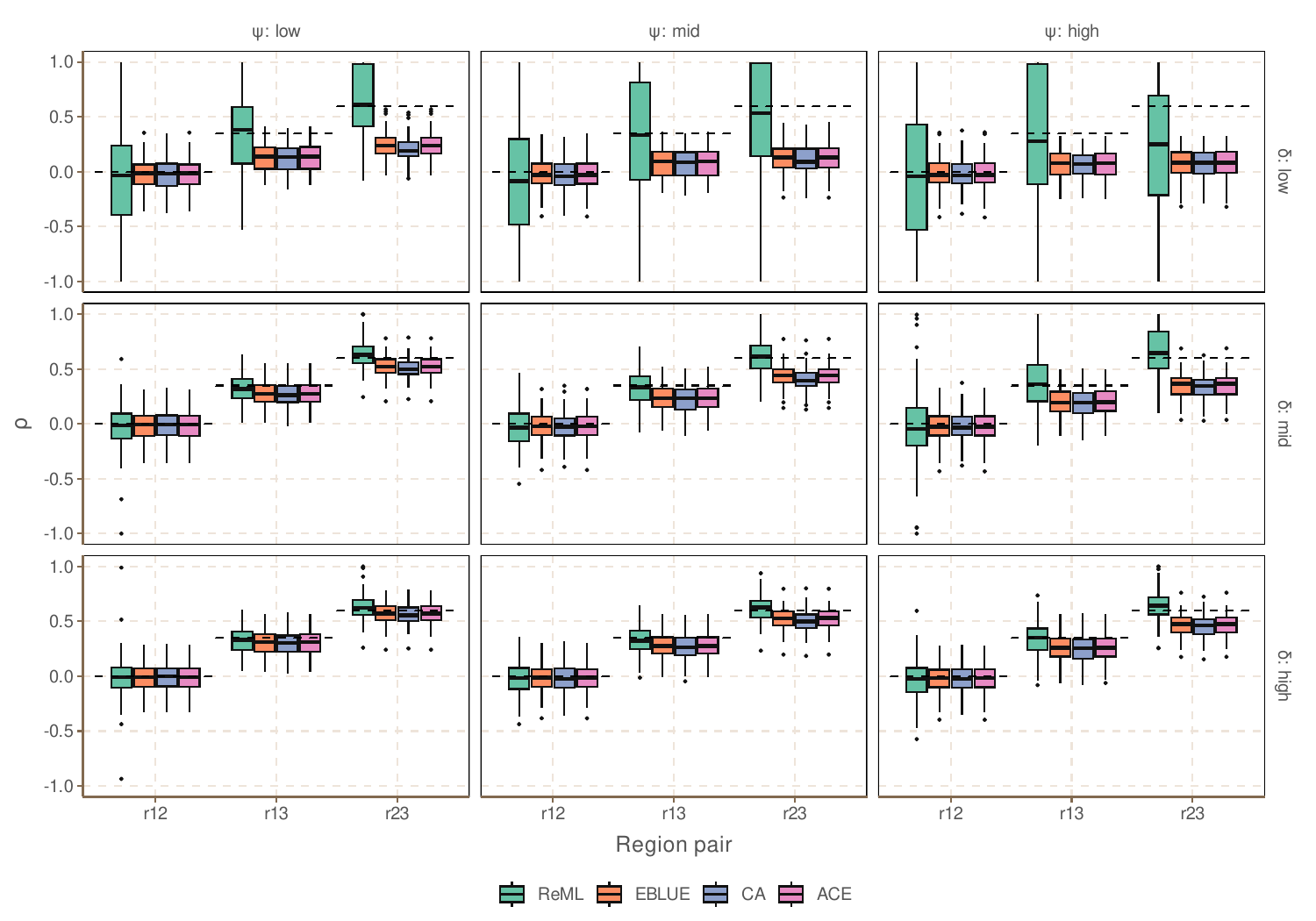}
 \caption{Distribution of $\hrReml_{jj'}$, $\hrEblue_{jj'}$, $\hrCA_{jj'}$, and $\hrACE_{jj'}$ for three region pairs over nine simulation scenarios under a $\operatorname{fgn}(0.9)$ temporal covariance structure
with $100$ replications in each scenario. Rows indicate low ($\delta = 0.1$), medium ($\delta = 0.5$), and high ($\delta = 0.7$) signal strengths while columns indicate low ($\psi = 0.2$), medium ($\psi = 0.5$), and high ($\psi = 0.8)$ intra-regional spatial correlations. The true correlations ($\rho_{12} = 0$, $\rho_{13} = 0.35$, $\rho_{23} = 0.6$) are marked by a dashed line. }
\label{fig:fgn}
\end{figure}

To generate the data under these alternative temporal covariance structures, we replace the temporal covariance kernel $H$ in \eqref{eq:latCovModel} with $Q_\mathrm{ar}$ or $Q_\mathrm{fgn}$
and then apply a discrete wavelet transform using a Daubechies filter of length $8$. The length of the pre-filtered time series is such that the filtering results in $60$ wavelet coefficients.
The results are shown in Web~Figures~\ref{fig:ar2} and \ref{fig:fgn}. We see that a misspecified temporal covariance is challenging for all methods in the low signal ($\delta = 0.1$) setting while higher spatial covariance ($\psi = 0.8$) increases the spread for any signal setting. Nevertheless, $\hrReml_{jj'}$ maintains good performance relative to $\hrCA_{jj'}$ when the signal strength increases.

We also study the behavior under a version of the nonstationary, locally anisotropic covariance kernel that was studied in \citet{castruccio2018scalable}.
Specifically, we will utilize the covariance kernel implied by Equation (3) in \citet{castruccio2018scalable}, with the number of components set to $L = 2$.
We construct this alternative covariance matrix as follows.
Let $\mathcal{V}_j$ be the collection of voxel coordinates for region $j$, in which there are $L_j$ voxels.
We partition $\mathcal{V}_j = \mathcal{V}_{j1} \cup \mathcal{V}_{j2}$ along an axis such that the partitions are roughly equal in size. Denote the centroid of $\mathcal{V}_{js}$ by $v^\ast_{js}$, $s = 1,2$.
Let $R_{js}$, $s = 1,2$, be diagonal matrices with positive values on the diagonal, which distort the Euclidean distance and lead to anisotropy when $R_{js}$ is not the identity matrix.

\begin{figure}[htbp!]
  \centering
  \includegraphics[width=\linewidth]{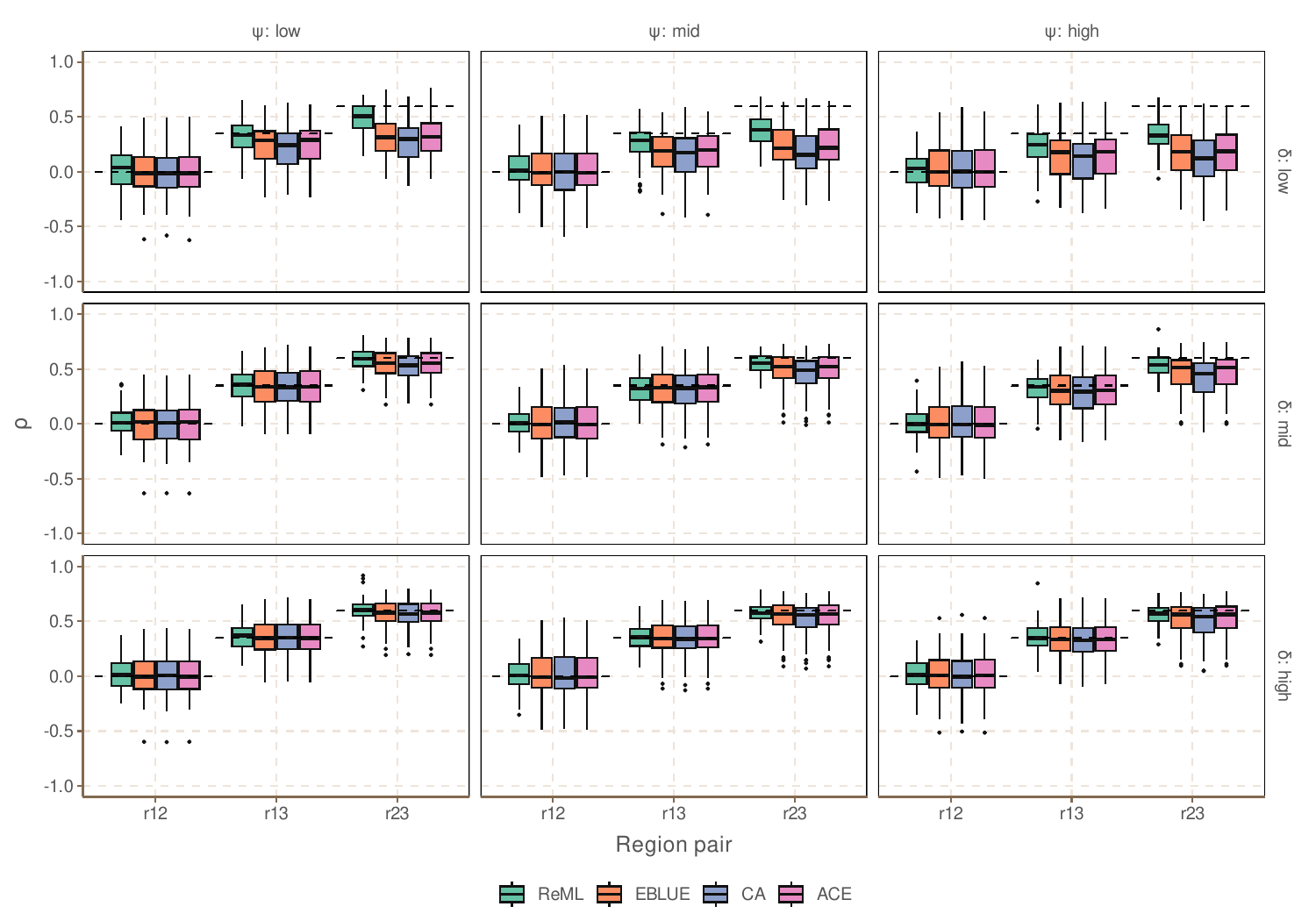}
 \caption{Distribution of $\hrReml_{jj'}$, $\hrEblue_{jj'}$, $\hrCA_{jj'}$, and $\hrACE_{jj'}$ for three region pairs over nine simulation scenarios under an anisotropic spatial covariance structure \citep{castruccio2018scalable}
with $100$ replications in each scenario. Rows indicate low ($\delta = 0.1$), medium ($\delta = 0.5$), and high ($\delta = 0.7$) signal strengths while columns indicate low ($\psi = 0.2$), medium ($\psi = 0.5$), and high ($\psi = 0.8)$ intra-regional spatial correlations. The true correlations ($\rho_{12} = 0$, $\rho_{13} = 0.35$, $\rho_{23} = 0.6$) are marked by a dashed line. }
 \label{fig:anisotropic}
\end{figure}

Then the spatial covariance matrix that replaces the matrix $C_j$ defined in \eqref{eq:latCovModel} is
\[
(\tilde C_j)_{ll'} = \sum_{s = 1}^2 w_{sl}w_{sl'}K(\lVert R_s(v_{jl} - v_{jl'}) \rVert; \nu_{js}, \phi_{js}),
\]
where $w_{sl} = \left(\lVert v_{jl} - v^*_{js} \rVert\right)^{-1}$ is the inverse distance from voxel $v_{jl}$ to the centroid of subregion $s$ and $K(\cdot; \nu, \phi)$
is the Mat\`ern covariance function with smoothness $\nu$ and scale $\phi$.
We set $R_{js} = \operatorname{diag}(1, 1.2, 1.5)$ and $\nu_{js} = 5/2$ for $s = 1,2$, while $\phi_{js}$ is determined through the spatial covariance simulation setting $\psi_j$ as usual.
The results are shown in Web~Figure~\ref{fig:anisotropic}. As expected, higher voxel-level spatial covariance under misspecification leads to difficulty for our method. However the performance improves as the signal strength increases.

\begin{figure}[htbp!]
  \centering
  \includegraphics[width=\linewidth]{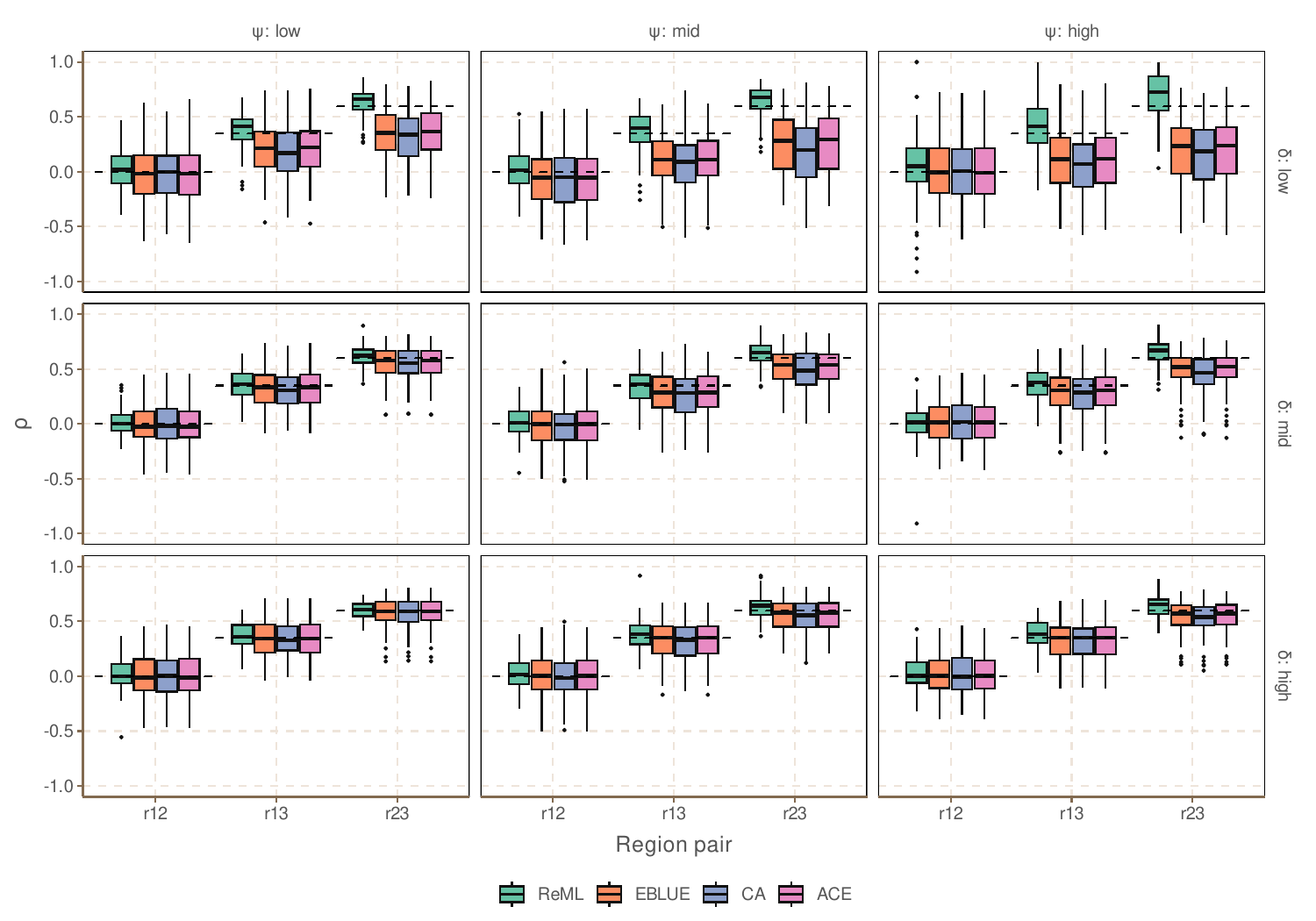}
 \caption{Distribution of $\hrReml_{jj'}$, $\hrEblue_{jj'}$, $\hrCA_{jj'}$, and $\hrACE_{jj'}$ for three region pairs over nine simulation scenarios under a non-separable spatio-temporal covariance kernel \citep{ip_li_nonsep}
with $100$ replications in each scenario. Rows indicate low ($\delta = 0.1$), medium ($\delta = 0.5$), and high ($\delta = 0.7$) signal strengths while columns indicate low ($\psi = 0.2$), medium ($\psi = 0.5$), and high ($\psi = 0.8)$ intra-regional spatial correlations. The true correlations ($\rho_{12} = 0$, $\rho_{13} = 0.35$, $\rho_{23} = 0.6$) are marked by a dashed line. }
 \label{fig:nonsep_matern}
\end{figure}

A final setting of interest is one where the spatio-temporal covariance is not separable between the spatial and temporal dimensions. To this end, we generate data from the stationary, non-separable covariance kernel described in \cite{ip_li_nonsep} given by
\[
M_{\nu,\alpha,\beta}(\vec h, u)=\frac{2^{1-\nu+\frac{d+1}{2}}}{\Gamma\left(\nu-\frac{d+1}{2}\right)}\left(\sqrt{\alpha^2 \vec h^2+\beta^2 u^2}\right)^{\nu-\frac{d+1}{2}} \mathcal{K}_{\nu-\frac{d+1}{2}}\left(\sqrt{\alpha^2 \vec h^2+\beta^2 u^2}\right)
\]
where $\vec h = \lVert \vec v - \vec v' \rVert$, $u = |t - t'|$, $d$ is the number of spatial dimensions, $\mathcal{K}(\cdot)$ is the modified Bessel function of the second kind, $\nu$ is a smoothness parameter, and $\alpha$ and $\beta$ control the relative scaling.
This non-separable kernel has marginally Mat\'ern spatial and temporal covariance.
For our simulations, we specify $d=3$, $\nu = 9/2$, $\alpha = \phi_\gamma$, with $\beta = 0.5$, which is analogous to the spatial and temporal scaling in the correctly specified setting.
The results are shown in Web Figure~\ref{fig:nonsep_matern}.
In all scenarios, $\hrReml_{jj'}$ has better performance over competitors with lower spread, especially in low signal settings.

\subsection{Robustness to stage-wise estimation}\label{sec:compare_oracle}
\begin{figure}[htbp!]
  \centering
  \includegraphics[width=\linewidth]{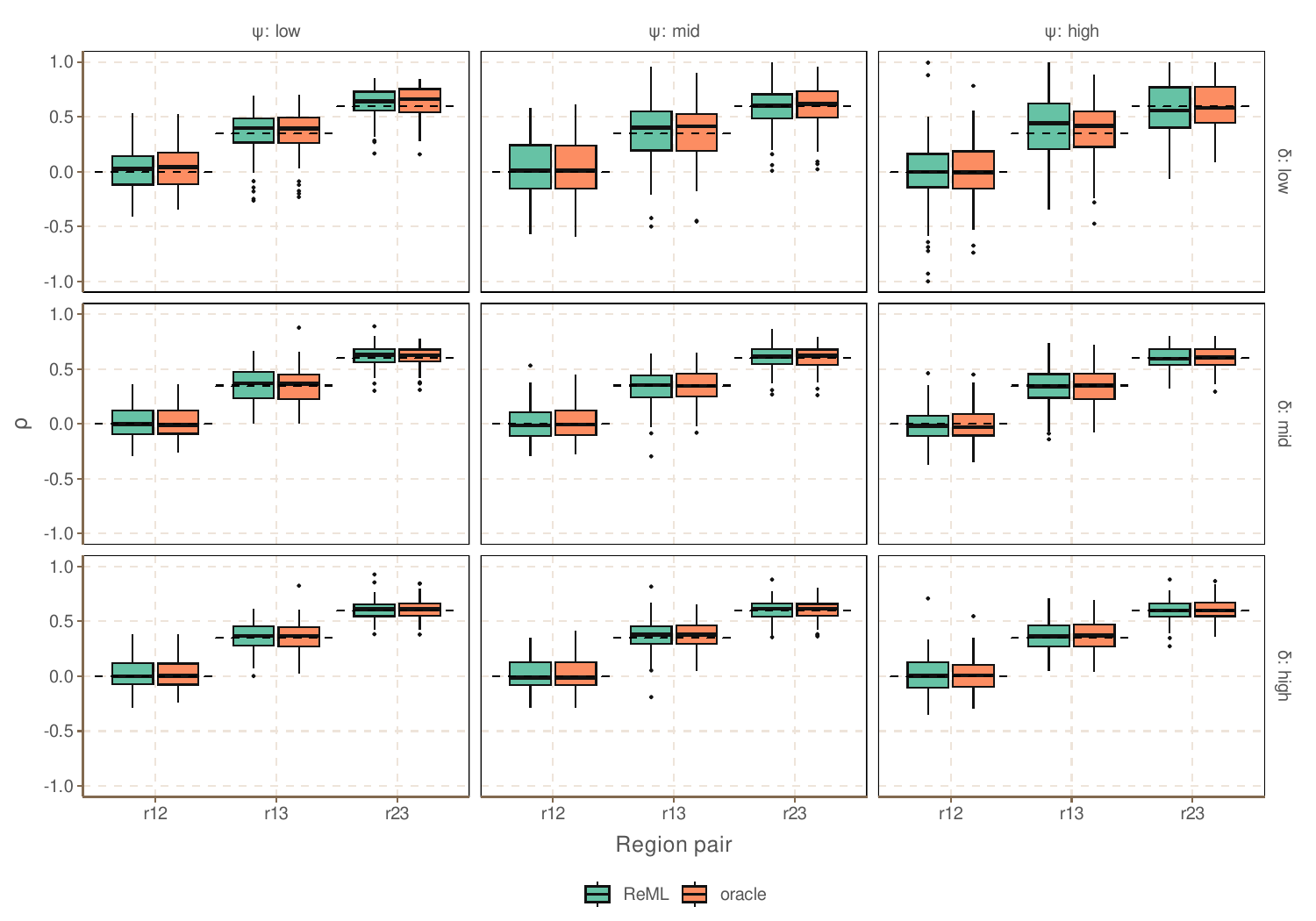}
 \caption{Comparison of $\hrReml_{jj'}$ with $\hrOracle_{jj'}$, which is achieved by running Stage 2 of our method with the true Stage 1 parameters plugged in and fixed. The performance is nearly identical in all simulation settings, with the largest differences in the difficult low $\delta$ settings.}
 \label{fig:oracle}
\end{figure}

Our method involves a two-stage plug-in procedure where Stage 1 parameters are plugged in and fixed throughout Stage 2. It is important to understand how estimation error in Stage 1 affects the final estimate of $\hrReml_{jj'}$ in Stage 2. In this section, we compare $\hrReml_{jj'}$ with $\hrOracle_{jj'}$, an \emph{oracle} estimator  where the true Stage 1 parameters are plugged in and fixed throughout Stage 2. The results are shown in Web Figure~\ref{fig:oracle} and we see that $\hrReml_{jj'}$ is close to the oracle estimator in all simulation settings. The largest differences occur in the low $\delta$ settings, where both methods exhibit higher spread.

\section*{Web Appendix B\\Discussion of noisy and noiseless models}\label{sec:noisyvnoiseless}
\begin{figure}[htbp!]
  \centering
  \includegraphics[width=\linewidth]{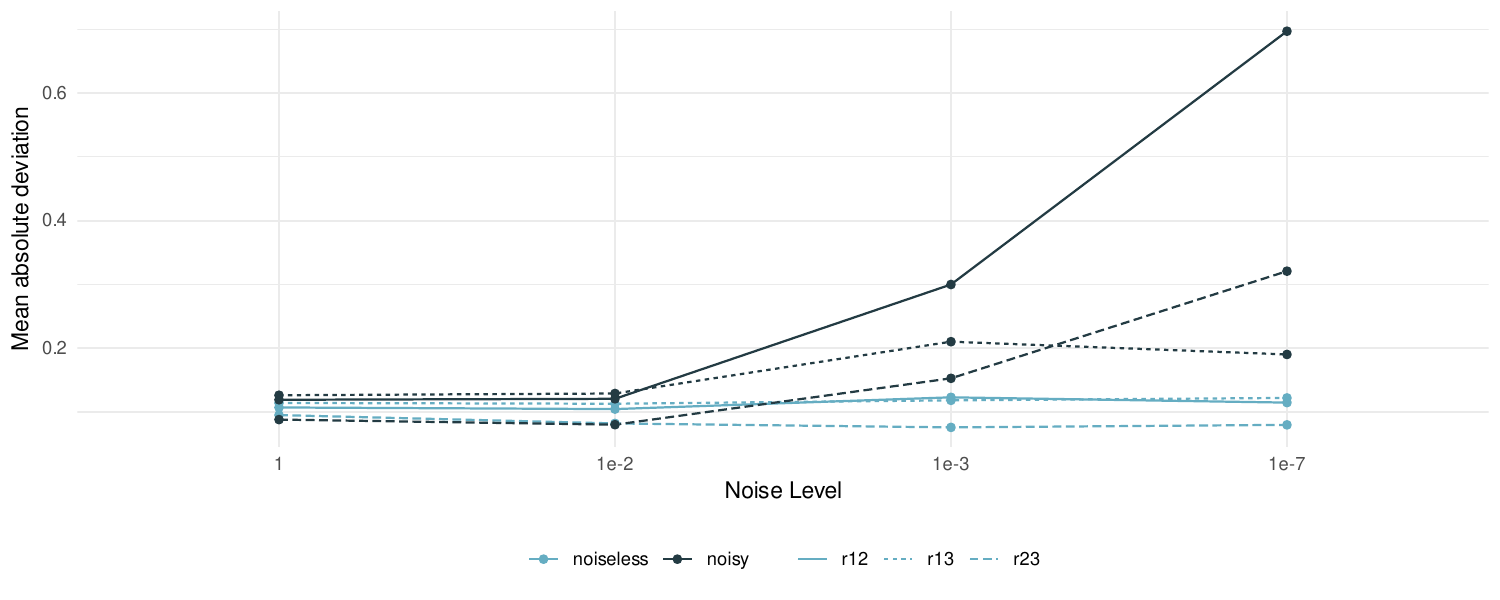}
 \caption{Comparison of noisy and noiseless models under different noise variance levels. For the mid signal, mid spatial covariance simulation setting, the mean absolute deviation is plotted for all region pairs. We see that as the true noise variance is small, the noisy model has difficulty identifying the variance components along with the i.i.d. noise. On the other hand, the noiseless model maintains reasonable performance even for extremely low noise variance levels.}
 \label{fig:noisy_vs_noiseless}
\end{figure}
We choose the noiseless model over the noisy model when the voxel-level variance dominates the overall noise variance as determined by inspecting the Stage 1 parameters.
This is motivated by Web~Figure~\ref{fig:noisy_vs_noiseless} which shows the mean absolute deviation of the noisy versus the noiseless estimates for different levels of $\sigma^2_\epsilon$ along on the $x$-axis, for the mid $\delta$ and mid $\psi$ setting. We see that the performance of the noisy model deteriorates as the overall noise variance vanishes by orders of magnitude. When $\sigma^2_\epsilon = 10^{-7}$, the noisy model struggles as the noise variance is significantly smaller than the variance of the latent signal.
In practice, we run the Stage 1 noisy model multiple times with different initializations. By inspecting the Stage 1 parameters, we detect if the region is in a low noise or high noise regime and we refit the Stage 1 with the noiseless model in the former case.
In our real data analysis, we choose the noiseless model for a region $j$ whenever $\log(k_{\gamma_j} + \sigma^2_{\gamma_j}) > 5$ or $\psi_j > 0.5$ as these indicate a strongly dominant latent signal over the noise variance.

\setcounter{section}{3}
\setcounter{subsection}{0}
\section*{Web Appendix C\\Discussion of Vecchia's approximation}\label{sec:supp_vecchia}
For datasets such as the HCP database, the large covariance matrices prohibit direct optimization of the ReML criteria for Stage 2 of our proposed method. In such cases, we can use Vecchia's approximation \citep{vecchia_estimation_1988} to get a computationally tractable optimization method while maintaining estimation accuracy.

Let $X_1$ and $X_2$ denote the $ML_1$ and $ML_2$ spatio-temporal points in regions 1 and 2. The accuracy of Vecchia's approximation depends on an ordering of the index set of all observations $\{1, \dotsc, M(L_1 + L_2)\}$ as well the choice of conditioning sets $\mathcal{J}_i$.
We use the \emph{maximum minimum distance} (MMD) ordering where points are sequentially chosen to have the maximum minimum distance to previously selected points.
We choose conditioning sets based on the nearest neighbor approach in \citet{guinness_permutation_2018} while accounting for the regional structure of our model.
Specifically, for a point $i$, $\mathcal{J}_i$ consists of the $100$ closest points to $i$ that precede $i$ in the MMD ordering, with the additional constraint that half the points come from each region. We term this approach \emph{region-aware nearest neighbors}. In all distance calculations, we scale down the spatial coordinates by a factor of $10$ so that difference between spatial and temporal dimensions are on the same scale.

\begin{figure}[htbb!]
  \centering
  \includegraphics[width=0.6\linewidth]{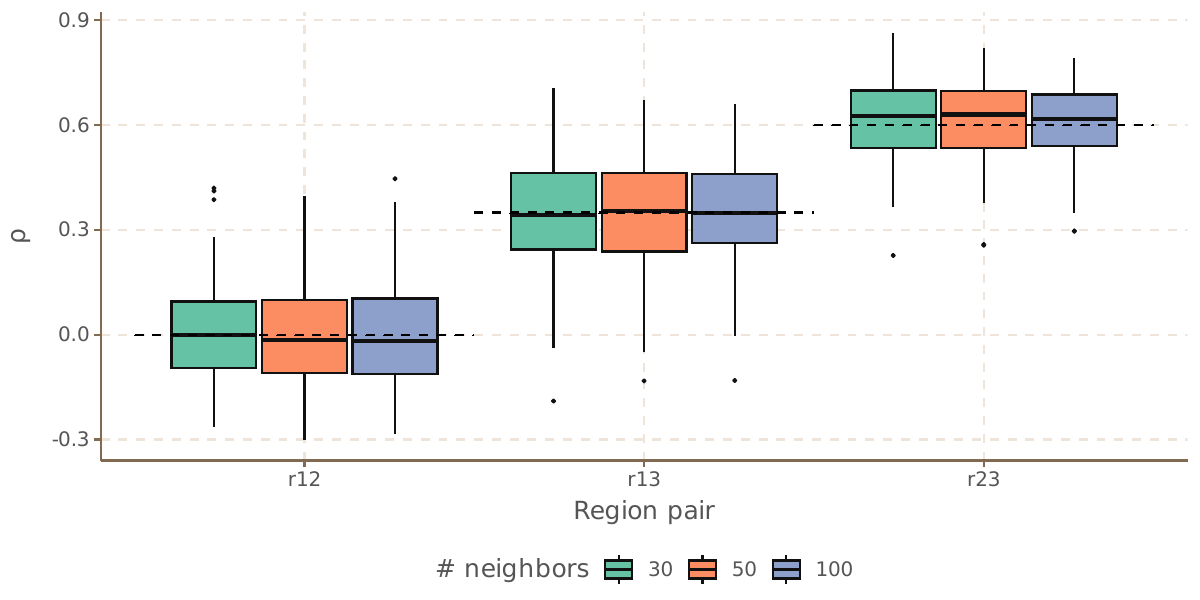}
 \caption{Distribution of $\hrVecchia_{jj'}$ for conditioning set sizes of $\abs{\mathcal{J}_i} = 30, 50, 100$. Results are over $100$ simulation runs over the $\delta = 0.5$ and $\psi = 0.5$ setting.}
 \label{fig:v_neighbors}
\end{figure}

Larger conditioning set sizes $\mathcal{J}_i$ leads to more accurate approximations at the cost of larger covariance matrix components in \eqref{eq:vecchia_approx}.
We chose $\abs{\mathcal{J}_i} = 100$ neighbors to balance between accuracy and our computational availability. It is recommended in \citet{guinness_permutation_2018} to choose from $30$ up to $100$ neighbors, although this is application dependent. Web Figure~\ref{fig:v_neighbors} shows that $\hrVecchia_{jj'}$ is fairly robust to the choice of neighbors beyond $30$, so reducing the conditioning set size is viable if computational resources are scarce.

\subsection{Results on HCP regions}
\label{sec:hcp-sim}
\begin{figure}[htbb!]
  \centering
  \includegraphics[width=\linewidth]{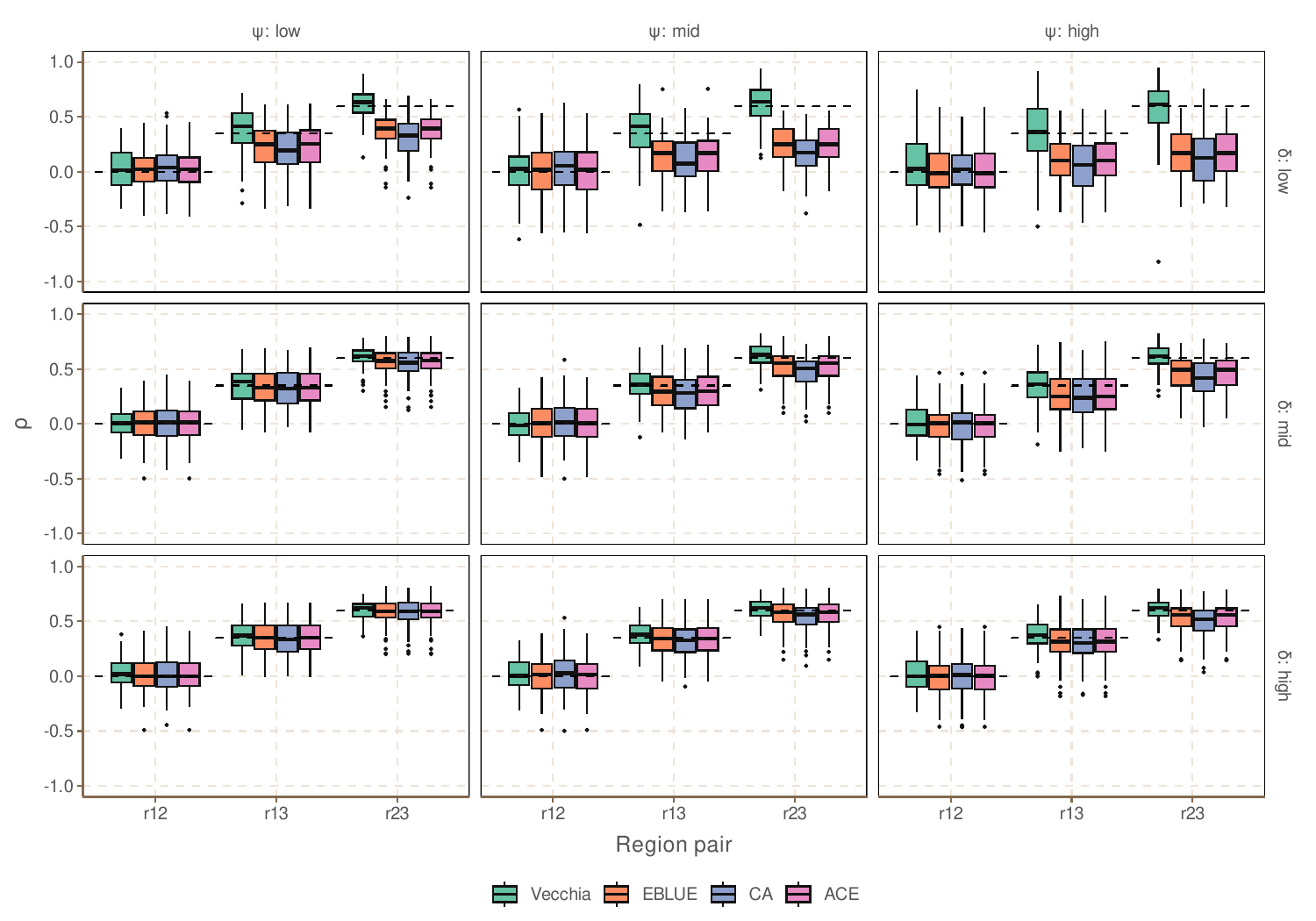}
 \caption{Distribution of $\hrVecchia_{jj'}$, $\hrEblue_{jj'}$, $\hrCA_{jj'}$, and $\hrACE_{jj'}$ over $100$ simulations and three levels of inter-regional correlation. The simulation settings described in Figure~\ref{fig:correctly} are used with a set of larger regions containing $L_1 = 166$, $L_2 = 233$, and $L_3 = 326$ voxels. Results show that the Vecchia's approximation estimator is not systematically biased in HCP voxel sets.}
 \label{fig:hcp_sim}
\end{figure}

It is important to check that Vecchia's approximation is not biased in spatial regimes that resemble those of the HCP test subjects studied in Section~\ref{sec:data}. We ran the simulation specification of Section~\ref{sec:sim_settings} using a set of regions containing $L_1 = 166$, $L_2 = 233$, and $L_3 = 326$ voxels whose spatial coordinates come from an arbitrarily selected HCP test subject. These regions are the first, second, and third quartiles in terms of number of voxels of the subject. The results shown in Web Figure~\ref{fig:hcp_sim} suggest that $\hrVecchia_{jj'}$ maintains unbiased performance in larger regions with HCP spatial structure.

\subsection{Timing on HCP region simulations}
\label{sec:hcp_sim_timing}
Computational cost and timing is an important practical consideration in applications.
Web Table~\ref{tbl:vec_sim_timing} shows the wall clock time to compute $\hrVecchia_{jj'}$ for the simulation study in Web Appendix~\ref{sec:hcp-sim}. We ran our method on a machine with 2.40GHz Intel Xeon processors. The quantity $M(L_j + L_{j'})$ is the total number of observations going into Vecchia's approximation. In our implementation, the Vecchia components in \eqref{eq:vecchia_approx} were split across five cores in each run.

\begin{table}[ht]
\centering
\begin{tabular}{cc|c}
\hline
$(L_j, L_{j'})$ & $M(L_j + L_{j'})$ & Runtime (s) \\
\hline
$(166, 233)$ & $23{,}940$      &  $96.8$ $(57.6)$          \\
$(166, 326)$ & $29{,}520$      &  $130.$ $(66.2)$           \\
$(233, 326)$ & $33{,}540$      &  $159.$ $(67.0)$ \\
\hline
\end{tabular}
\caption{\label{tbl:vec_sim_timing}Mean (and standard deviation) of runtime in seconds to compute $\hrVecchia_{jj'}$ over $100$ simulations each in nine simulation settings and three region pairs. $(L_j, L_{j'})$ are the count of voxels in each region pair and $M=60$ time points were used. The spatial coordinates come from an arbitrarily selected HCP test subject.}
\end{table}

\subsection{Timing on HCP data}
\begin{figure}[htbp!]
  \centering
  \includegraphics[width=0.7\linewidth]{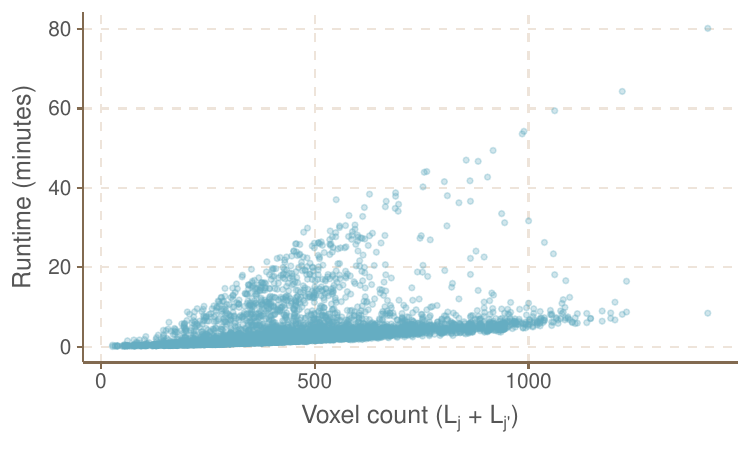}
 \caption{Time to compute $\hrVecchia_{jj'}$ for an arbitrarily selected HCP subject. Each point represents a region pair $(j, j')$, with $\binom{92}{2}$ pairs in all, and the runtime in minutes is plotted against the total number of voxels $L_j + L_{j'}$.}
 \label{fig:runtime_subject}
\end{figure}

In Web Figure~\ref{fig:runtime_subject} we plot the runtime to compute $\hrVecchia_{jj'}$ for an arbitrarily selected subject from our HCP data analysis in Section~\ref{sec:data}.
The plot shows all pairs of $92$ default mode regions studied in our analysis. The regions range from $27$ to $1419$ voxels and the median region size is $428$ voxels.
With parallelizing the Vecchia components across five cores, the median time to compute $\hrVecchia_{jj'}$ was $127$ seconds. We see that a small number of cases failed to converge quickly to a solution, yet the runtime for these cases appears linear in the number of voxels.

\setcounter{section}{4}
\setcounter{subsection}{0}
\section*{Web Appendix D\\Asymptotic inference}\label{sec:confint}
\subsection{Asymptotic development}
We extend the discussion in Section~\ref{sec:asymp} to develop the asymptotic distribution of the ReML estimator $\hat{\bmom}^{\mathrm{ReML}}$ of the full parameter vector $\bmom$ defined in \eqref{eq:mdlInterReML}.
As mentioned in the main article, the asymptotic regime is that of a diverging number $M$ of time points or wavelet coefficients.
\citet{cressie1993asymptotic} developed the general asymptotic theory for ReML estimators in linear mixed models which \citet{cressie1996asymptotics} then applied to spatial regression settings similar to those considered by \citet{mardia1984maximum}, providing practically verifiable sufficient conditions.

Consider the case of $J = 2$ regions for \eqref{eq:mdlXvec}, so that the sample size is $N=M(L_1+L_2)$.
Denote the negative restricted log-likelihood by $\Lr(\bmom)$. Let $\bmV^{(i)}(\bmom) = \partial \bmV(\bmom) / \partial \omega_i$, where $\omega_i$ is the $i$-th element of $\bmom \in \mathbb{R}^p$.
Denote the $p\times p$ matrix of second-order partial derivatives of $\Lr(\bmom)$ by
\[
\bmmcI_N(\bmom) = \left\{\frac{\partial^2\Lr(\bmom)}{\partial\omega_i \partial \omega_j} \right\}_{i,j=1}^p.
\]
The elements of the Fisher information matrix are given by
\begin{equation*}
    \label{eq:iMat_both}
\begin{split}
    \mathbb{E}_{\bmom} (\bmmcI_N(\bmom))_{ij} = \frac{1}{2}\Tr\left\{ \bmH(\bmom) \bmV^{(i)}(\bmom) \bmH(\bmom) \bmV^{(j)}(\bmom)\right\}
\end{split}
\end{equation*}
for $1 \leq i, j \leq p$.
Under the regularity conditions of \citet{cressie1996asymptotics}, the full ReML estimator $\hat{\bmom}^{\mathrm{ReML}}$ is asymptotically normal:
\begin{equation}
    \label{eq:fisher_info_vecchia}
    \left\{\E{\bmmcI_N(\bmom)}\right\}^{1/2} (\hat{\bmom}^{\mathrm{ReML}} -\bmom ) \rightsquigarrow \mathbf{N}(\bmZero, \bmI_p)
\end{equation}
with $\rightsquigarrow$ denoting convergence in distribution.

We validate the asymptotic normality of $\hrReml_{jj'}$ and $\hrVecchia_{jj'}$ by constructing confidence intervals.
Let $\hat\rho$ stand for either $\hrReml_{jj'}$ or $\hrVecchia_{jj'}$ and let $Z_{\hat\rho} = f(\hat\rho)$ where $f(r) = \arctanh(r)$ is the Fisher $Z$-transformation.
The delta method gives the following $1-\alpha$ confidence interval on the $Z$-transform scale:
\[
\left(\operatorname{lower}(Z_{\hat\rho},\alpha),\, \operatorname{upper}(Z_{\hat\rho},\alpha)\right)
=
\left({Z}_{\hat\rho} - z_{\alpha/2}f'(\hat\rho)\sqrt{\Var{\hat{\rho}}},\,
Z_{\hat\rho} + z_{\alpha/2}f'(\hat\rho)\sqrt{\Var{\hat{\rho}}}
\right),
\]
where $\Var{\hat\rho}$ comes from (19) and $z_{\cdot} = \Phi^{-1}(1-\cdot)$ denotes the quantile function of the standard normal distribution.
Mapping back to the correlation scale yields the $1-\alpha$ confidence interval for $\hat\rho$:
\begin{equation*}
\label{eq:hrReml_{jj'}CI}
C_{1-\alpha}(\hat\rho) = \left(f^{-1}(\operatorname{lower}(Z_{\hat\rho},\alpha)),\, f^{-1}(\operatorname{upper}(Z_{\hat\rho},\alpha))\right).
\end{equation*}
For $\hrCA_{jj'}$ we use a standard approach, constructing intervals on the $Z$ scale using the standard error $1/\sqrt{N-3}$ and transforming back to the correlation scale.

\subsection{Simulation study of coverage}

\begin{figure}[htbp!]
  \centering
  \includegraphics[width=\linewidth]{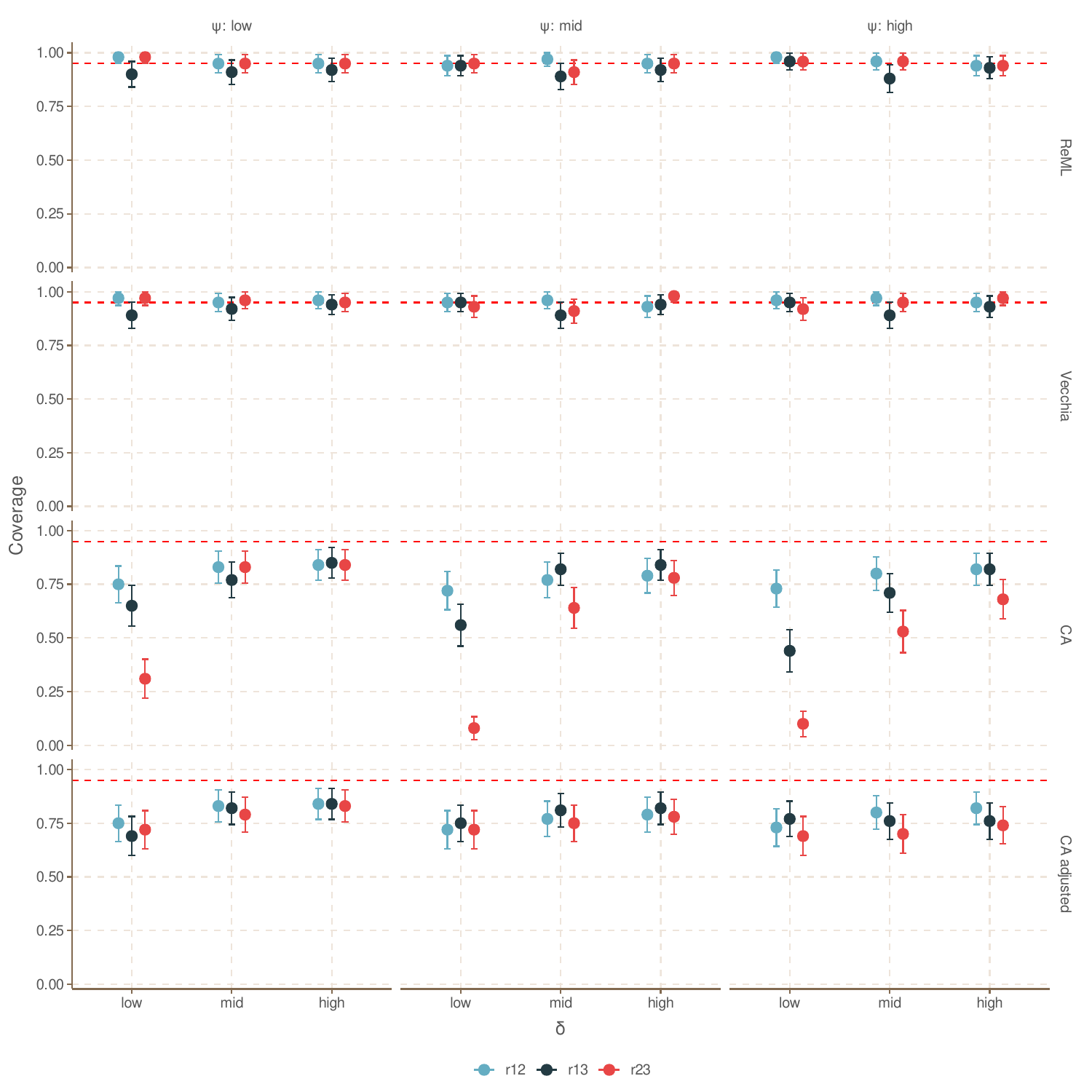}
 \caption{Coverage of $95\%$ asymptotic confidence intervals for $\hrReml_{jj'}$, $\hrVecchia_{jj'}$, $\hrCA_{jj'}$, and adjusted $\hrCA_{jj'}$ over $100$ simulations and three levels of inter-regional correlation.
 The CA interval is constructed from standard methods while adjusted CA refers to scaling $\hrCA_{jj'}$ by the denominator of \eqref{eq:rcaLimit1} so that the target estimand is $\rho_{jj'}$.
 The $y$-axis is the proportion of simulations where the $95\%$ confidence interval contains the true parameter, where $\rho_{12} = 0$, $\rho_{13} = 0.35$, and $\rho_{23} = 0.6$, shown as Bernoulli trials. The simulation settings described in Figure~\ref{fig:correctly} are used.}
 \label{fig:all_95_ci}
\end{figure}

Web Figure~\ref{fig:all_95_ci} supplements Figure~\ref{fig:vecchia_95_ci} by showing the $95\%$ confidence interval coverage of ReML and (unadjusted) CA in addition to that of Vecchia and adjusted CA. We see that the coverage of $\hrReml_{jj'}$ and $\hrVecchia_{jj'}$ are similar and that that of $\hrCA_{jj'}$ is poor due to CA targeting a parameter corrupted by noise. The adjusted CA rescales the estimand to be the desired $\rho_{jj'}$ yet the intervals still systematically under-cover.

\subsection{FDR thresholding for HCP data}

\begin{figure}[htbp!]
  \centering
  \includegraphics[width=\linewidth]{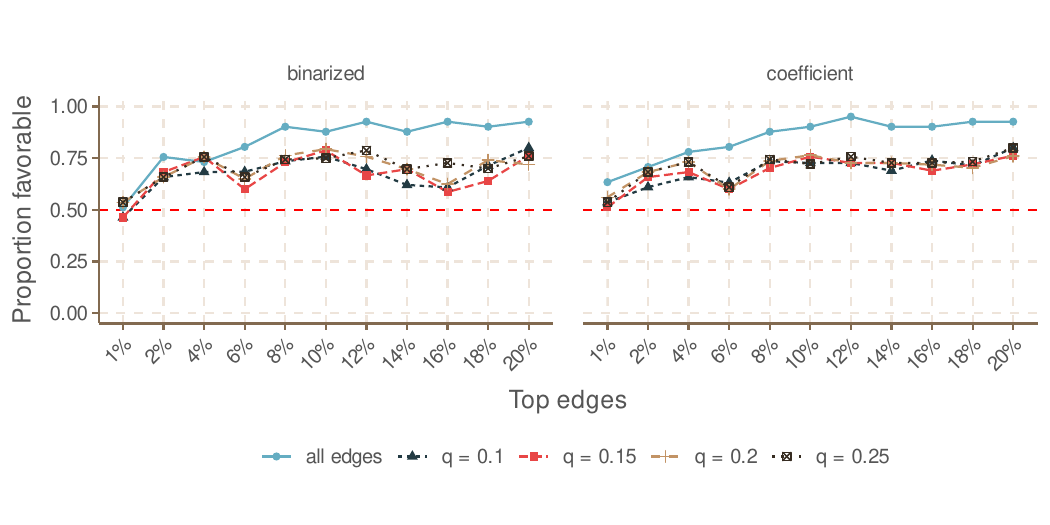}
 \caption{Proportion of the $42$ HCP test-retest subjects with higher concordance correlation coefficient (CCC) for networks constructed from our proposed mixed model relative to those from the Correlation of Averages. The $y$-axis shows the proportion favorable to the mixed model, with the dashed red line marking the $50\%$ reference point (no difference between methods). Results are shown across percentages of top edges used ($x$-axis), comparing graphs constructed from all edges (circles, solid line) versus statistically significant edges only (dashed lines with Benjamini--Yekutieli FDR control, $q < 0.1$, $0.15$, $0.2$, and $0.25$). Panels are faceted by whether graphs were binarized (left) or retained correlation coefficients (right).}
 \label{fig:cccprop_vq}
\end{figure}

In Section~\ref{sec:data} we investigate the CCC of HCP subjects in a test-retest study after thresholding for each subject both the estimated test and retest networks by significance and additionally imposing a Benjamini-Yekuteily FDR adjustment with $q < 0.2$. In Web Figure~\ref{fig:cccprop_vq} we plot the proportion of HCP subjects with CCC favorable to our proposed mixed model for different thresholds $q < q_0$ for $q_0 \in \{0.1, 0.15, 0.2, 0.25\}$. This demonstrates that the favorable performance is not sensitive to the choice of the threshold.

\label{lastpage}

\end{document}